\documentclass[preprint,letter,11pt,numberedappendix]{emulateapj}
\usepackage{epsf}
\usepackage{apjfonts}
\usepackage{graphicx}

\newcommand{\oi}{[O\,{\sc i}]}
\newcommand{\oii}{[O\,{\sc ii}]}
\newcommand{\oiii}{[O\,{\sc iii}]}

\newcommand{\Ni}{[N\,{\sc i}]}
\newcommand{\nii}{[N\,{\sc ii}]}

\newcommand{\sii}{[S\,{\sc ii}]}
\newcommand{\siii}{[S\,{\sc iii}]}

\newcommand{\hei}{He\,{\sc i}}

\newcommand{\neiii}{[Ne\,{\sc iii}]}

\newcommand{\ariii}{[Ar\,{\sc iii}]}
\newcommand{\ariv}{[Ar\,{\sc iv}]}

\newcommand{\ha}{H$\alpha$}
\newcommand{\hb}{H$\beta$}

\newcommand{\te}{$T_{\rm e}$}
\newcommand{\Ne}{$n_{\rm e}$}

\received{}
\revised{}
\accepted{}

\slugcomment{}
\shorttitle{The detection of C$_{60}$ in the PN M1-11}
\shortauthors{Otsuka et al.}

\begin{document}

\title{
The detection of C60 in the well-characterized planetary nebula M1-11
}

\author{Masaaki Otsuka\altaffilmark{1,2}, 
F.~Kemper\altaffilmark{1}, 
S.~Hyung\altaffilmark{3}, 
B.~A.~Sargent\altaffilmark{2,4}, 
M.~Meixner\altaffilmark{2},
A.~Tajitsu\altaffilmark{5}, 
K.~Yanagisawa\altaffilmark{6}}

\affil{
$^{1}$Institute of Astronomy and Astrophysics, Academia Sinica
P.O. Box 23-141, Taipei 10617, Taiwan, R.O.C.; otsuka@asiaa.sinica.edu.tw}
\affil{$^{2}$Space Telescope Science Institute, 3700 San Martin Drive,
Baltimore, MD 21218, U.S.A.}
\affil{
$^{3}$School of Science Education (Astronomy), Chungbuk
National University, CheongJu, Chungbuk
361-763, Korea}
\affil{
$^{4}$Center for Imaging Science and Laboratory for Multiwavelength Astrophysics, Rochester Institute of Technology, 54 Lomb Memorial Drive, Rochester, NY 14623, USA}
\affil{
$^{5}$Subaru Telescope, NAOJ, 650 North A'ohoku Place, Hilo,
Hawaii 96720, U.S.A.}
\affil{
$^{6}$Okayama Astrophysical Observatory (OAO), NAOJ, Kamogata, Okayama 719-0232, Japan
}

\begin{abstract}
  We performed multiwavelength observations of the young planetary
  nebula (PN) M1-11 and obtained its elemental abundances, dust mass, and
  the evolutionary status of the central star. 
The \emph{AKARI}/IRC, VLT/VISIR, and \emph{Spitzer}/IRS
  spectra show features due to carbon-rich dust, such as the 3.3, 8.6,
 and 11.3 $\mu$m features due to polycyclic aromatic hydrocarbons
 (PAHs), a smooth continuum attributable to amorphous carbon, and the
 broad 11.5 and 30 $\mu$m features often ascribed to SiC and MgS, respectively.
We also report the  presence of an unidentified broad feature at 16--22 $\mu$m, similar to the feature
  found in Magellanic Cloud PNe with either C-rich or
  O-rich gas-phase compositions. 
We identify for the first time in M1-11 spectral lines at 8.5 (blended with
 PAH), 17.3, and 18.9 $\mu$m that we attribute to the C$_{60}$
 fullerene. 
This identification is strengthened by the fact that other
  Galactic PNe in which fullerenes are detected, have
  similar central stars, similar gas-phase abundances, and a similar
  dust composition to M1-11. The weak radiation field due to the
  relatively cool central stars in these PNe may provide
  favorable conditions for fullerenes to survive in the circumstellar
  medium. 
  Using the photo-ionization code
  {\sc Cloudy}, combined with a modified blackbody, we have fitted the
  $\sim$0.1--90 $\mu$m spectral energy distribution and determined the
  dust mass in the nebula to be $\sim$3.5$\times$10$^{-4}$ $M_{\odot}$. 
Our chemical abundance analysis and SED model 
 suggest that M1-11 is perhaps a C-rich PN with C/O ratio in the
  gas-phase of +0.19 dex, and that it evolved from a 1--1.5
  $M_{\odot}$ star. 
\end{abstract}

\section{Introduction}

The most  stable fullerene is C$_{60}$ (Draine 2011). Fullerenes
together with other carbon dust such as graphite are expected to be
important components of the interstellar medium (ISM), because they 
contribute to interstellar extinction. For example, Dopita et
al. (1997) argued the possibility that the deep 2200 {\AA} absorption feature
in the low-excitation planetary nebula (PN) SMP LMC8 might be by a surface
charge slop resonance on C$_{60}$. The graphite grains have a spectral peak
around this wavelength due to $\pi$$\rightarrow$$\pi^{\ast}$ electron excitations. Fullerene
and polycyclic aromatic hydrocarbons (PAHs) resemble graphite; therefore
such grains have strong electronic transitions around 2200 {\AA} (Draine
2011). The investigation of circumstellar carbon grains such as 
C$_{60}$ would be important to understand ISM evolution more deeply.

Recent $Spitzer$/IRS studies show that fullerenes C$_{60}$ are
detected in several young PNe, a proto-PN, and two post asymptotic giant branch
stars (post AGB) in the Milky Way (Cami et al. 2010;
Garc\'{i}a-Hern\'{a}ndez et al. 2010; Zhang \& Kwok 2011; Gielen et
al. 2011) and also in a handful young PNe in the Magellanic Clouds
(Garc\'{i}a-Hern\'{a}ndez et al. 2011a; Bernard-Salas et al. 2012). 
The detection cases are increasing, however the excitation and 
formation mechanism is still unclear (e.g., Cami et al. 2011; 
Garc\'{i}a-Hern\'{a}ndez 2012). The detections generally 
suggest that fullerenes can 
survive or be observed in a C-rich environment with a weak radiation field.

M1-11 (PN G232.8$-$04.7) is a good sample in C$_{60}$ formation in
a circumstellar environment, 
because most of the C$_{60}$ PNe have cool central stars ($\sim$30,000
K) and C-rich nebulae, and they are very young ($\sim$ 1000
yr). Indeed, M1-11 has the cool central star (29\,300 K, Phillips 2003), and the
nebula is relatively compact, with a size of $\lesssim$6$''$ in diameter 
in H$\alpha$ emission, and it appears to be a very young object
($\sim$1000 yr after leaving the AGB phase, this paper). 
Henry et al. (2010) measured the C abundance in M1-11
using recombination lines (RLs) and the O abundance with collisionally
excited lines (CELs, number density ratio of C/O = 79.4). To date, only upper limits to the intensities of
the CELs C~{\sc iii}] 1906 and 1909 {\AA} are obtained (Kingsburgh \&
Barlow 1994), and thus the gas phase C/O ratio of M1-11 derived from the same type
of emission line is still unknown. However, M1-11 is also known to be 
a dust-rich PN, showing predominantly C-rich dust. Silicon carbide (SiC) and amorphous silicate features are seen in the
$IRAS$/LRS data (Zhang \& Kwok 1990). A significant near-infrared
excess suggests the presence of hot dust (Phillips \& Ramos-Larios 2005;
Zhang \& Kwok 1990), although there is also a contribution from the
3.3 $\mu$m emission feature due to PAHs in the near infrared (Allen et al.~1982). 
Longer wavelength
emission features at 6.2, 6.9, 7.7, and 8.6 $\mu$m due to PAHs are
reported by Cohen et al. (1986). A very weak and tentative feature
is seen around 6.9-7.0 $\mu$m (see Cohen et al. 1986). Indeed, there are other C$_{60}$ transitions at 7.0 and 8.5
$\mu$m but Cohen et al. (1986) show some evidence (although the quality
of their data is not satisfactory) for a possible detection also in M1-11.
Therefore, M1-11 is perhaps a C-rich PN.

To confirm whether C$_{60}$ in M1-11 is real and also obtain insights about
C$_{60}$ formation, we need to investigate the physical and chemical
properties of the dust and ionized nebula and the nature of the central
star, then we need to compare the derived quantities with those in
C$_{60}$ PNe. To characterize M1-11, we obtained continuous data
from the UV to the far-infrared using several instruments, and we comprehensively investigated this PN.

In this paper, we discuss C$_{60}$ in M1-11 based on the information of the
dust and gas compositions and the evolutionary status of the central
star. In
section 2, we describe the UV- to mid-infrared (mid-IR) spectroscopic 
data from the International Ultraviolet Explorer (\emph{IUE}),
Subaru/HDS, OAO/ISLE, \emph{AKARI}/IRC, VLT/VISIR, and \emph{Spitzer}/IRS, as
well as narrow-band imaging obtained with WFPC2 on the Hubble Space
Telescope (\emph{HST}). The description 
in Section 2.6 includes a list of dust features seen in M1-11,
specifically, we report the discovery of weak features at 8.5, 17.3 and
18.9 $\mu$m lines, which are attributed to the C$_{60}$ fullerenes. 
The derivation of the ionic and elemental
abundances in the ionized nebula is given in Section 3. In Section 4,
we discuss the observed C and O gas abundances and compare them with
the predictions from nucleosynthesis models for AGB stars. Using the
photo-ionization code {\sc Cloudy} (Ferland et al. 1998), we fitted the
spectral energy distribution (SED) and determined the dust mass and the
evolutionary status of the central star. Section 4 also includes a
discussion on the formation of C$_{60}$, and a comparison of the
physical properties of M1-11 to those of other Galactic PNe that exhibit
fullerene features. A summary and future prospects are given in Section
5.

\section{Observations and Data reduction}

\subsection{Subaru/HDS observations}

We obtained optical spectra of M1-11 using the High-Dispersion Spectrograph
(HDS; Noguchi et al. 2002) attached to one of the two Nasmyth foci of
the 8.2-m Subaru telescope, on October 6, 2008 (program ID: S08B-110,
PI: M.Otsuka) and October 18, 2005 (PI: A.Tajitsu). The spectra were
taken in two wavelength ranges: 3600--5400 {\AA} (the blue spectra,
taken in 2008) and 4600--7500 {\AA} (the red spectra, taken in 2005).

When we obtained the blue spectra, an atmospheric dispersion corrector
(ADC) was used to minimize the differential atmospheric dispersion
over the broad wavelength region. We used a slit width of
$1.\hspace{-2pt}''2$ (0.6 mm) and a 2$\times$2 on-chip binning.  We
set the slit length to be $8''$ (4.0 mm), which fitted the nebula well
and allowed us to directly subtract sky background from the object
frames. The slit position angle (P.A.) was $\sim$225$^{\circ}$.  The
CCD sampling pitch along the slit length projected on the sky was
$\sim0.\hspace{-2pt}''276$ per binned pixel.  The resolving power
reached around $R$ $>$33\,000, which is derived from the mean of the
full width at half maximum (FWHM) of narrow Th-Ar and night sky
lines. The total exposure time was 600 sec (=300 sec $\times$2
frames).  For the flux calibration, blaze function correction, and
airmass correction, we observed G192B2B as a standard star.

For the red spectra, we used the red image de-rotator and set it to
P.A.=90$^{\circ}$.  We set the slit width to $0.\hspace{-2pt}''6$ and
the slit length to 7$''$ and selected a 1 (wavelength
dispersion)$\times$2 (spatial direction) on-chip binning. The
resulting spectral resolution $R$ is $>$65\,000. We used an exposure
time of 300 sec and observed G192B2B as a standard star.

For both sets of observations, we took several bias,
instrumental flat lamp, and Th-Ar comparison lamp frames.
We are interested in detecting weak C, N, and O recombination lines. 
The peak intensities of these lines are typically
$\sim$10-20$\%$ higher than the local continuum, and therefore a
high signal-to-noise ratio (S/N) of the continuum is necessary. The
resulting S/N  after subtraction of the sky background was found to range
from $\sim$5 at $\sim$3700 {\AA} to $\sim$30 at $\sim$5200 {\AA} in the
blue spectrum, and from $\sim$5 at $\sim$4800 {\AA} to 15 at $\sim$6700
{\AA} in the red one.

Data reduction of the Subaru/HDS spectra and analysis of the emission 
lines was done with the
long-slit reduction package {\tt noao.twodspec} available in
IRAF\footnote[7]{IRAF is distributed by the National Optical Astronomy
  Observatories, which are operated by the Association of Universities
  for Research in Astronomy (AURA), Inc., under a cooperative
  agreement with the National Science Foundation.}, and was performed
in the same manner as described by Otsuka et al. (2010). When
measuring the fluxes of the emission-lines, we assumed that the line
profiles were Gaussian and we applied a multiple Gaussian fitting
technique.

The line fluxes were de-reddened using
\begin{equation}
\label{redc}
\log_{10}\left[\frac{I(\lambda)}{I{\rm (H\beta)}}\right] =
\log_{10}\left[\frac{F(\lambda)}{F{\rm (H\beta)}}\right] + c({\rm H\beta})f(\lambda),
\end{equation}
\noindent

where $I(\lambda)$ and $F(\lambda)$ are the de-reddened and the
observed fluxes at $\lambda$, respectively, and $f(\lambda)$ is the
interstellar extinction parameter at $\lambda$, from the
reddening law of Cardelli et al. (1989) with $R_{V}$ = 3.1. The
interstellar reddening correction was performed using the reddening
coefficient $c$({\hb}), at {\hb}.  We compared the observed Balmer
line ratios of H$\gamma$ (blue spectrum) or {\ha} (red spectrum) with
{\hb} to the theoretical ratio computed by Storey \& Hummer (1995)
assuming the electron temperature {\te} = 10$^{4}$ K and the electron
density {\Ne} = 10$^{4}$ cm$^{-3}$ assuming that the nebula is
optically thick in Ly-$\alpha$ (Case B of Baker \& Menzel 1938). We
derived $c$({\hb}) = 1.677$\pm$0.008 for the blue and
1.218$\pm$0.017 for the red spectra.

The flux scaling was performed using all emission lines detected in
the overlap region between the blue and the red spectra. The
de-reddened relative intensities $I$($\lambda$) detected in both
spectra are consistent within 10$\%$ of each other. The combined
de-reddened spectrum is presented in Fig.~\ref{hds}, and the detected
lines are listed in the Appendix (Table \ref{hdst}).  We have detected
over 160 emission lines, thus exceeding the number of detections by Henry et
al.~(2010), who report more than 70 lines in the 3700--9600 {\AA}
spectra. Our measurements of the line intensities $I$($\lambda$) are in 
agreement with the results from Henry et
al.~(2010) within a $\sim$14 $\%$ error.

Specifically, we detected C~{\sc ii}, N~{\sc ii}, and O~{\sc ii} RLs, 
and highly excited lines due to He~{\sc ii},
C~{\sc iii}, and N~{\sc iii}. These high excitation lines show a
relatively broad FWHM ($\sim$0.8-1.6 {\AA}) compared to typical
nebular lines ($\sim$0.2--0.5 {\AA}).  It is possible that the He~{\sc
  ii}, C~{\sc iii}, and N~{\sc iii} lines are not of nebular origin
but of stellar origin, because the effective temperature of the
central star (29\,300 K; Phillips 2003) is not high enough for species
with an ionization potential (IP) $\gtrsim$40 eV to exist in the
nebula. For example, we did not detect any nebular lines from species
with an IP $\gtrsim$40 eV, such as {\neiii} $\lambda\lambda$3876/3967
(IP$>$41 eV) and {\ariv} $\lambda$4711/40 ($>$ 40.7 eV). The IPs of
He~{\sc ii}, C~{\sc iii}, and N~{\sc iii} are 54.4, 47.9, and 47.5 eV,
respectively.

For our analysis, we also used the emission-line
fluxes in the 7700--9300 {\AA} range measured by Henry et al. (2010) 
scaled in such a way that the shorter wavelength part of their
spectrum matches our HDS observations.

\begin{figure}
\epsscale{1.1} 
\plotone{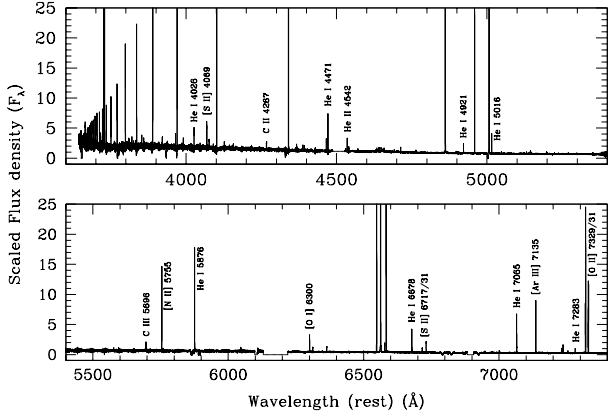} \caption{The scaled and de-reddened HDS
spectrum of M1-11. The wavelength is shifted to the rest wavelength in
air.\label{hds}}
\end{figure}

\subsection{\emph{HST}/WFPC2 archive data}
We downloaded archival \emph{HST}/WFPC2 photometry in the F656N (6564
{\AA}/28 {\AA}) filter (P.I.: R.Sahai; PID: 8345), which
traces the {\ha} emission.  We reduced the
photometric data using the standard \emph{HST} pipeline with
MultiDrizzle, and present the drizzled M1-11 F656N image in
Fig.~\ref{image}. The plate scale is 0.025$\arcsec$ pixel$^{-1}$.  The
image shows that M1-11 is an elongated nebula; the dimensions of the
bright rim are $\sim$0.8$''$ along P.A.=--27$^{\circ}$ and
$\sim$0.5$''$ along 
P.A.=+63$^{\circ}$.

\subsubsection{The total H$\alpha$ and H$\beta$ fluxes}
The de-reddening formula (eq.~(1)) requires the total {\hb} flux over
the entire nebula to obtain the line fluxes in the  
\emph{AKARI} and \emph{Spitzer} spectra. This can be derived from the
WFPC2 {\ha} image. 
We find that the total flux in the F656N filter is 1.77(--11) $\pm$ 3.48(--13) erg s$^{-1}$ cm$^{-2}$ integrated over the entire PN (we will use the notation $X(-Y)$ for $X\times10^{-Y}$,
hereafter), where we assume that the uncertainty corresponds to the
standard deviation of the background. Using the HDS red spectrum and
the transmission curve of the F656N filter, we estimated $\sim$12.5
$\%$ of the total measured flux to be due to a local continuum and the
{\nii} 6548 {\AA} line flux. Thus, we estimated the solo H$\alpha$
line flux to be 1.55(--11) erg s$^{-1}$ cm$^{-2}$. Using the observed
$F$({\ha})/$F$({\hb}) ratio (6.57) in the HDS red spectrum, we
derive the total $\log$$F$({\hb}) to be --11.629 erg s$^{-1}$
cm$^{-2}$, which is comparable to $\log$$F$({\hb})=--11.84 erg
s$^{-1}$ cm$^{-2}$ measured by Cahn et al. (1992).

\subsubsection{The hydrogen density profile}
\begin{figure}
\epsscale{1.1}
\plotone{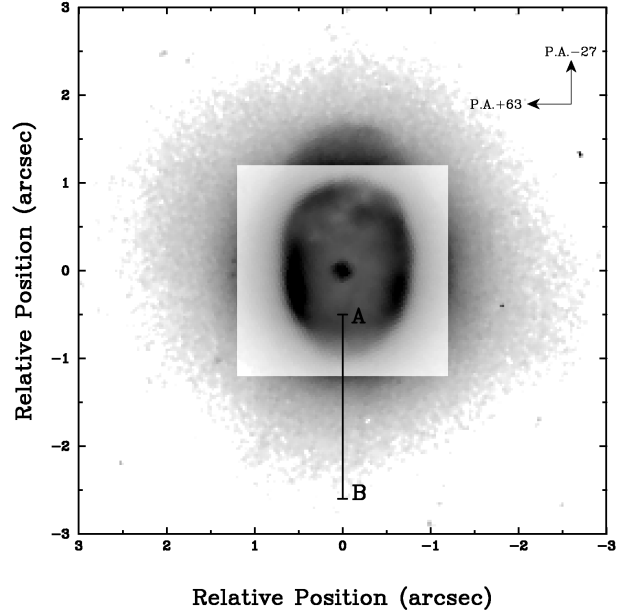}
\caption{\emph{HST}/WFPC2 F656N image of M1-11, rotated according to
  elongation of the nebula.  In the inner 2$''$$\times$2$''$ box, the
  gray scale is adjusted to show the brightness of the bright rim and
  the central star. The radial profiles in {\ha} and $n$(H$^{+}$) from
  A (Dec.~relative position=--0.5$''$) towards B (--2.7$''$) are
  presented in Fig.~\ref{radial}(a) and (b), respectively.}
\label{image}
\end{figure}

\begin{figure}
\epsscale{1.1}
\plotone{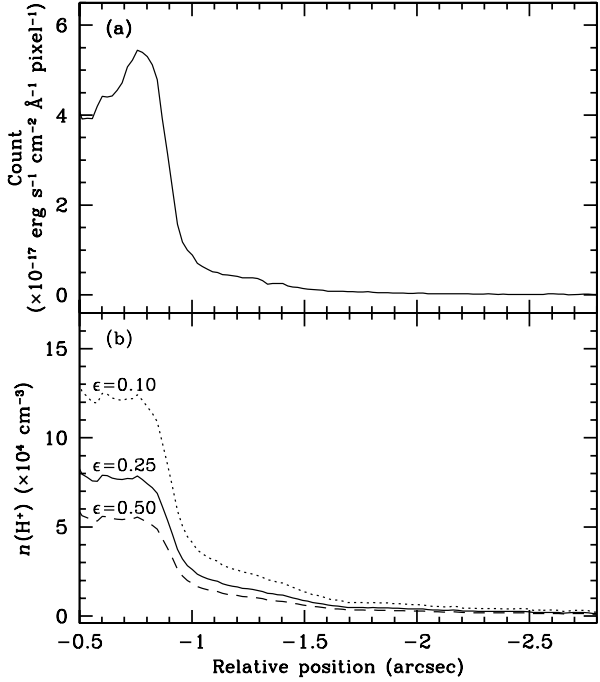}
\caption{The radial profiles from A to B (Fig.~\ref{image}) in {\ha}
  ({\it upper}) and $n$(H$^{+}$) ({\it lower}) with different filling
  factors $\epsilon$.}
\label{radial}
\end{figure}

In Fig.~\ref{radial}(a), we present the radial profile from A toward B
indicated in Fig.~\ref{image}.  Based on this radial profile, we
examined the ionized hydrogen density $n$(H$^{+}$) as a function of
the distance from the central star $R$. When we restrict the
integration to the optional portion of the nebula, the de-reddened
observed {\ha} flux using $c$({\hb})=1.218 and the reddening law of
Cardelli et al. (1989) with $R_{V}$ = 3.1, $I_{l}$(\ha) in erg
s$^{-1}$ cm$^{-2}$ is given by:
\begin{equation}
4\pi D^2 I_{l}({\rm H}\alpha) = \int 4{\pi}j({\rm H}\alpha){\epsilon}~dV_{l},
\end{equation}
\noindent
where $D$ is the distance to M1-11 from us; $j$(H$\alpha$) is the
emission coefficient; and  $\epsilon$ is the filling factor which is the
fraction of the nebular volume filled by ionized gas. $V_{l}$ is the
volume of the of the nebula. In Case B,
\begin{equation}
\frac{4{\pi}j({\rm H}\alpha)}{n({\rm H}^{+})n_{\rm e}} \simeq 3.856\times10^{-25}\left(\frac{10^{4}}{T_{\rm e}}\right)^{-1.077}.
\end{equation}
with $T_{\rm e}$ = 5400 K derived from the Balmer jump (See section
3.2), $D$ = 2.1 kpc (Tajitsu \& Tamura 1998),
and assuming $n_{\rm e}$ $\simeq$ $n$(H$^{+}$), $n$(H$^{+}$) can be written
as a function of the distance from the central star $R$ in cm as
\begin{equation}
n({\rm H}^{+}) \simeq 3.74\times10^{19}\left(\frac{I_{l}({\rm H}\alpha)}{\epsilon R}\right)^{0.5}.
\label{density}
\end{equation}
The resulting $n$(H$^{+}$) profile from A towards B with different
values for $\epsilon$ is presented in Fig.~\ref{radial}(b).  If we
assume that the ionized gas is concentrated within $R$=1.0$''$ and
that the density has a constant value of 10$^{5}$ cm$^{-3}$ obtained
from Balmer decrements, then we find that $\epsilon$ is around 0.2. We
used these $n$(H$^{+}$) profiles with different values for $\epsilon$
in the SED modeling (See section 4.2).


\subsection{ISLE Observations}
\begin{figure}
\centering
\epsscale{1.1}
\plotone{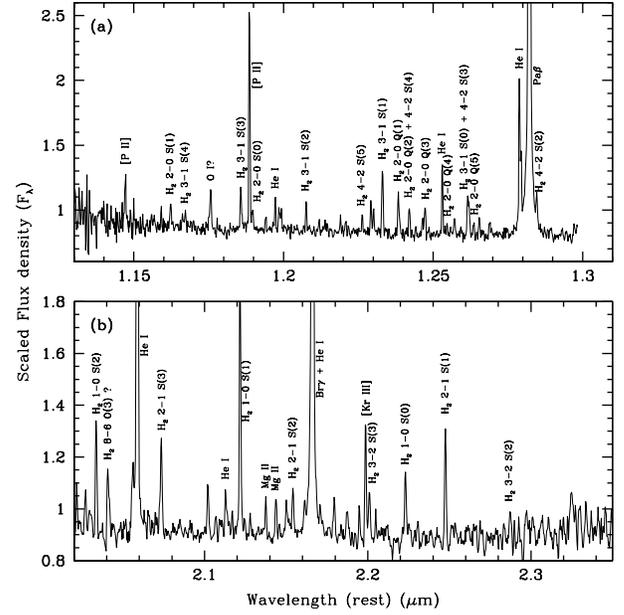}
\caption{The ISLE $JKs$-band spectra of M1-11.}
\label{isle}
\end{figure}

We obtained $J$ and $Ks$ band medium-resolution ($R$$\sim$2500)
spectra using the near-infrared imager and spectrograph ISLE
(Yanagisawa et al. 2006, 2008) attached to the Cassegrain focus of the
1.88-m telescope at the Okayama Astrophysical Observatory. The
observations were done in ISLE engineering time in March 2008 ($Ks$)
and January 2010 ($J$). The detector of ISLE is a 1K$\times$1K HgCdTe
HAWAII array. We used a science grade detector for the $J$-band
observations and an engineering grade detector for the $Ks$-band
observations.  The entrance slit width was 1$''$ for both sets of
observations.  We fixed the P.A.~at 90$^{\circ}$ during the
observations.  The sampling pitches in wavelength were
$\sim$1.68$\times$10$^{-4}$ and $\sim$3.4$\times$10$^{-4}$ $\mu$m
pixel$^{-1}$ in the $J$ and $Ks$ spectra, respectively, while the
sampling pitch in the space direction was 0$\farcs25$ pixel$^{-1}$ for
both spectra.  We observed standard stars HIP35132 (A0V) and HIP35180
(A1V) for the $J$-band and HIP31900 (F0V) for the $Ks$-band spectra at
different airmasses to calibrate the flux levels, and correct for
telluric absorption and airmass extinction.  We observed M1-11 in a
series of 120 sec exposures in both observing modes. The total
exposure times were 3600 sec for the $J$-band spectra and 3480 sec for
the $Ks$-band spectra, respectively.  Dark frames with the same
science exposure time, Ar and Xe lamp frames, and on- and off-dome
flat frames were also taken.  For further wavelength calibration and
distortion correction, OH-lines recorded in the object frames were
used. The data reduction was performed in a standard manner using IRAF. The interstellar reddening corrected spectra are presented
in Fig.\ref{isle}(a) and (b). For this correction, we adopted 
$c$({\hb}) applied in the HDS red spectrum. The resulting S/Ns are $>$40 in the
$J$-band and $>$30 in $Ks$-band spectra at the continuum level.

We detected more than 50 lines in these spectra, including a series of
vibration-rotation excited lines of molecular hydrogen (H$_{2}$), 
as listed in the Appendix (Table \ref{isle_spec}). The line fluxes were
normalized such that $I$(Pa$\beta$)=100 in the $J$-band and
$I$(Br$\gamma$)=100 in the $Ks$-band spectra.

\begin{figure}
\centering
\epsscale{1.1}
\plotone{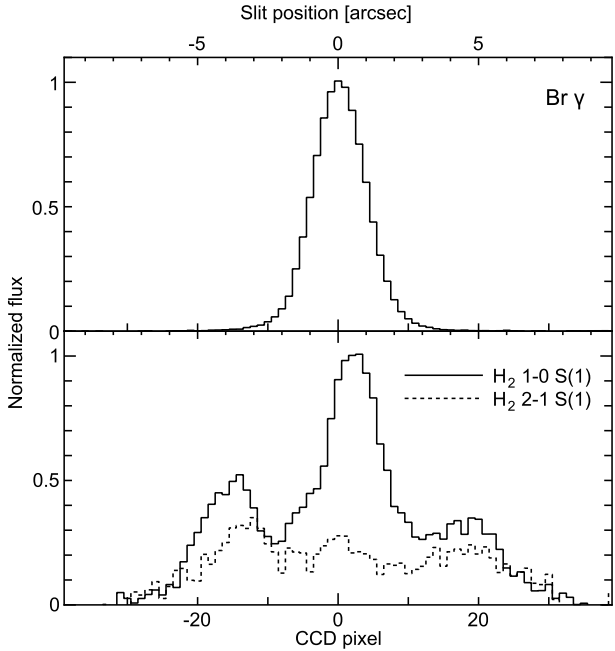}
\caption{Spatial profiles of Br $\gamma$ ({\it upper panel}), 
H$_{2} 1-0 S(1)$ and $2-1 S(1)$ lines ({\it lower panel}) 
along the slit in the $K$-band spectrum. After subtracted the 
continuum, profiles are normalized by the intensity peak of Br$\gamma$ 
and H$_{2} 1-0 S(1)$.}
\label{h2}
\end{figure}

Fig.\ref{h2} shows the spatial profiles of Br$\gamma$ and H$_{2}$ 1-0 
S(1)/2-1 S(1) lines in the $K$-band spectrum. In the both spectra, 
the H$_{2}$ lines are easily distinguished with other ionic lines 
by their spatial spread upto $\sim$ 16$''$ in the diameter. 
The ratio of H$_{2}$ 1-0 S(1)/2-1 S(1) is a traditional shock indicator 
(e.g, Hora \& Latter 1994, 1996; Kelly \& Hrivnak 2005). 
The ratio in M1-11 is $\sim$4.5 at the center of the nebula, and it
decreases up to $\sim$1.0 outside of the ionized region. 
The ratio of H$_{2}$ 1-0 S(1)/2-1 S(1) = 4.5 along the optical nebula 
($\lesssim$6$''$ in diameter) indicates a mix of UV and shock excitation 
(this is usual in proto-PNe, e.g., 
Kelly \& Hrivnak 2005). Furthermore, the detection of a
series of H$_{2}$ lines with upper vibrational level ($v\geq 3$) in
$J$-band indicates that these lines are excited by fluorescence through
the absorption of UV photons from the central star in photodissociation region (PDR).

\begin{figure}
\centering
\epsscale{1.1}
\plotone{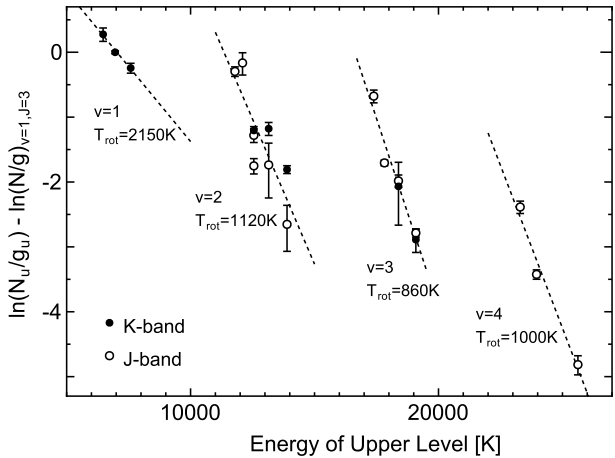}
\caption{
Molecular hydrogen excitation diagram from the full $J$- 
(open circles) and $K$-band (filled circles) spectra. 
Shown are the upper state vibration-rotation level populations 
relative to that in the $v=1$, $J=3$ level plotted against 
the energy of the upper state in Kelvin (K). $g$ is the statistical
 weight. An ortho-to-para ratio of 3 is assumed. The 
points within vibrational levels fall on separate lines, as expected fluorescent-excited 
emission. Linear fits to the data on each vibrational level 
are plotted with the derived rotational temperatures.
}
\label{ed}
\end{figure}

The excitation diagram of H$_{2}$ lines in the entire slit of the spectra
is shown in Fig.\ref{ed}. An ortho-to-para ratio of 3 is assumed. 
It clearly shows that the vibrational excitation temperature ($T_{vib}$) exceeds the rotation excitation 
temperature ($T_{rot}$), indicating fluorescence emission 
(cf. Shupe et al. 1998 for BD+30$^{\circ}$ 3639; Hora 
\& Latter 1994 for M2-9). The difference of the excitation temperature
between different rotation levels ($T_{rot}$$\sim$ 2150 K for $v=1$ and
$T_{rot}$$\sim$1000 K for $v\geq2$) indicates H$_{2}$ lines are
collisionally (shock) excited in a part (the center) of the
nebula. However,  it is evident that the UV excitation in PDR is still
dominant for the most part of H$_{2}$ emission in M1-11.

Any pure rotational lines in $Spitzer$/IRS are not detected (see Table \ref{akari_sp}).

The lines at 1.15 and 1.19 $\mu$m are identified with [P~{\sc
  ii}] (in the $^{3}P_{1}$--$^{1}D_{2}$ and
$^{3}P_{2}$--$^{1}D_{2}$ transitions, respectively), representing the
discovery of these lines in M1-11.  Adopting the
transition probabilities of Mendoza \& Zeippen (1982), the collisional
impacts of Tayal (2004a) and the level energy listed in Atomic Line
List v2.05b12\footnote[8]{see
  http://www.pa.uky.edu/$^{\sim}$peter/newpage/}, the expected [P~{\sc
  ii}] $I$(1.19 $\mu$m)/$I$(1.15 $\mu$m) ratio is 2.63 in $T_{\rm
  e}$=10\,000 K and $n_{\rm e}$=5$\times$10$^{4}$ cm$^{-3}$. The
observed line ratio (2.79$\pm$0.44) agrees well with the theoretical
value, which confirms the identification of the [P~{\sc ii}] 1.15/1.19 $\mu$m
lines. Our measurement of the $I$([Kr~{\sc iii}] 2.19
$\mu$m)/$I$(Br${\gamma}$) ratio of 3.44$\pm$0.23 also agrees with
Sterling \& Dinerstein (2008; 3.22$\pm$0.26).

\subsection{\emph{IUE} archival data}
The N$^{2+}$ abundance can be estimated from the N\,{\sc iii}]
$\lambda$1750 line, present in archival \emph{IUE} spectra which we
retrieved from the Multi-mission Archive at the STScI (MAST).  We
collected low-resolution \emph{IUE} spectra taken by the Short
Wavelength Prime (SWP) and Long Wavelength Prime (LWP) cameras (file
ID: SWP25846, LWP05896, and LWP05897), all of which were made using
the large aperture (10.3$\times$23 arcsec$^{2}$). In these spectra, we
identified the He~{\sc ii} 1640 {\AA} and N~{\sc iii}] 1750 {\AA}
lines.  We determined that $c$({\hb})=0.67$\pm$0.12, by comparing the
theoretical ratio of He\,{\sc ii}
$I$($\lambda$1640)/($\lambda$4686)=6.56 to the observed value, in the
case of $T_{\rm e}$=10$^{4}$ K and $n_{\rm e}$ = 10$^{4}$ cm$^{-3}$ as
given by Storey \& Hummer (1995). The interstellar extinction
correction was made using equation (1).  The flux
measurements of the detected lines along with the normalized
values are listed in Table \ref{iue_sp}. While we did not detect the C~{\sc iii}]
$\lambda\lambda$1906/09 {\AA} lines in the SWP and LWP spectra,
Kingsburgh \& Barlow (1994) show $I$(C~{\sc
  iii}]$\lambda\lambda$1906/09) to be 6 with an uncertainty greater 
than a factor of 2.

\begin{deluxetable}{@{}clccr@{}}
\tablecolumns{5}
\tablecaption{Lines detected in the \emph{IUE} spectra.}
\tablewidth{240pt}
\tablehead{
\colhead{$\lambda_{lab}$}&\colhead{Ion}&\colhead{$f$($\lambda$)}&\colhead{$F(\lambda)$$^{a}$}&\colhead{$I(\lambda)$}\\
\colhead{({\AA})}& \colhead{}& \colhead{}&\colhead{(erg s $^{-1}$ cm$^{-2}$)}&\colhead{($I$(H$\beta$)=100)}
}
\startdata
1640    & He~{\sc ii}     & 1.177 & 5.03(--14) $\pm$ 9.99(--15) & 0.51 $\pm$ 0.20 \\
1750    & N~{\sc iii}]    & 1.154 & 1.51(--13) $\pm$ 1.52(--14) & 1.48 $\pm$ 0.51 
\enddata
\tablenotetext{a}{$\log$$F$(H$\beta$) = --11.63 erg s$^{-1}$ cm$^{-2}$.}
\label{iue_sp}
\end{deluxetable}

\subsection{\emph{AKARI}/IRC archival data}
\begin{figure}
\centering
\epsscale{1.1}
\plotone{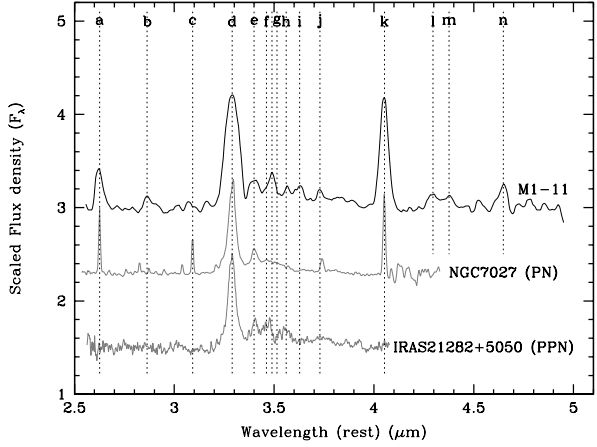}
\caption{Comparison of the \emph{AKARI}/IRC spectrum of M1-11 (upper line),  the archival
 $ISO$/SWS spectra of the PN NGC7027 and the PPN IRAS21282+5050 (middle
 \& lower). The identified lines are indicated by the the
 broken lines. The IDs are indicated by lower case letters;
a: Br$\beta$ 2.63 $\mu$m;
b: H~{\sc i} 5-11 2.86 $\mu$m;
c: PAH 3.29 $\mu$m C-H stretch + H~{\sc i} 5-9 3.29 $\mu$m;
d: PAH 3.38/3.40 $\mu$m asymmetric CH$_{3}$, CH$_{2}$ stretch;
e: PAH 3.46 $\mu$m lone C-H stretch;
f: PAH 3.49/3.51 $\mu$m symmetric CH$_{3}$, CH$_{2}$ stretch;
g: PAH 3.56$\mu$m aldehydes C-H stretch;
h: H~{\sc i} 6-19,20 3.63 $\mu$m;
j: Br$\alpha$ 4.05 $\mu$m;
k: H~{\sc i} 4-5,6-14 4.05 $\mu$m;
l: He~{\sc i} 3-5 4.30 $\mu$m;
m: H~{\sc i} 6-12 4.38 $\mu$m;
n: Pf$\beta$ 4.65
$\mu$m (Reference of the wavelength of the hydrocarbon lines: Kwok 2007). }
\label{akari}
\end{figure}

We analyzed the 2.5--5.5 $\mu$m prism spectra of M1-11 taken with
the Infrared Camera
spectrograph (IRC; Onaka et al. 2007) on board of the \emph{AKARI} satellite (Murakami
et al. 2007). The data were obtained as part of a mission
program, PNSPEC (data ID: 3460037, PI: T.Onaka), on April 11, 2009.
The used observing window was 1$'$$\times$1$'$.
For the data reduction, we used the IRC Spectroscopy Toolkit for the Phase 3
data version. 
Fig.~\ref{akari} shows the IRC spectrum with a local dust continuum
subtracted. The S/N is $>$30 for the dust continuum.
Several prominent lines are visible, and their central wavelengths are
indicated by dotted lines. The line fluxes
are listed in Table\ref{akari_sp}. For the IRC spectra, we derived that 
$c$({\hb})=1.40$\pm$0.03
by comparing the observed intensity ratios of H~{\sc i} 4--5 
(Br$\alpha$ 4.051 $\mu$m), 4--6 (Br$\beta$
2.625 $\mu$m), and 5--7 (Pf$\beta$ 4.653 $\mu$m)
to {\hb} and the theoretical values of Storey \& Hummer (1995) for the
case of 10$^{4}$ K and 10$^{4}$ cm$^{-3}$. To correct for interstellar
reddening, we used the ratio of the extinction at each wavelength to
the $B-V$ color excess, $A_{\lambda}$/$E(B-V)$, given by Fluks et al.
(1994), in combination with the correlation between {\hb} and the
color excess, $c$({\hb})=1.47$E(B-V)$, from Seaton (1979).

In the \emph{AKARI} spectra we also found an emission band at 3.2--3.6
$\mu$m, which may be due to aromatic and aliphatic hydrocarbon
species. A similar feature is seen in PN NGC7027 and proto-PN (PPN) 
IRAS21282+5050, the latter of which has a [WC11] central star. \emph{ISO}/SWS
archival spectra of NGC 7027 and IRAS 21282+5050 are shown
for reference in Fig.~\ref{akari}.
The resonance at 3.3 $\mu$m is attributed to vibrational transitions
in polycyclic aromatic hydrocarbons (PAHs; e.g.~Draine 2011). We also
recognize the 6.2, 7.7, and 11.3 $\mu$m resonances due to PAHs in the
spectra of M1-11. In particular, the 11.3 $\mu$m C-H out-of-plane
bending mode is seen in the \emph{Spitzer}/IRS spectrum (See next
section). Cohen et al. (1986) already detected the 6.2, 6.7, 7.7, and
8.6 $\mu$m resonances in M1-11. The 3.3 $\mu$m band profile of M1-11
is similar to the ones seen in NGC 7027 and IRAS 2182+5050, and thus
we assume that the emission seen in M1-11 is also due to PAHs. Indeed,
the 3.3 $\mu$m feature was first detected by Allen et al. (1982), who
also measured its flux.  Using the theoretical intensity ratio of
H~{\sc i} $I$(5--9) to $I$(5--11) = 1.86 in the case of $T_{\rm e}$=10$^{4}$ K and
$n_{\rm e}$=10$^{4}$ cm$^{-3}$, we removed the contribution from H~{\sc i}
$n$=5--9 to the 3.3 $\mu$m feature and estimated the flux due to PAHs
$I$(PAH 3.3 $\mu$m) to be 7.63(--12) erg s$^{-1}$ cm$^{-2}$, which is
about twice as large as the measurement by Allen et al.~(1982;
3.1(--12) erg s$^{-1}$ cm$^{-2}$).

\begin{deluxetable}{@{}clccr@{}}
\tablecolumns{5}
\tablecaption{Detected lines in the \emph{AKARI}/IRC,
 \emph{Spitzer}/IRS, and VLT/VISIR observations.}
\tablewidth{240pt}
\tablehead{
\colhead{$\lambda_{vac}$}&\colhead{Ion}&\colhead{$f$($\lambda$)}&\colhead{$F(\lambda)$$^{a}$}&\colhead{$I(\lambda)$}\\
\colhead{($\mu$m)}& \colhead{}& \colhead{}&\colhead{(erg s $^{-1}$ cm$^{-2}$)}&\colhead{($I$(H$\beta$)=100)}
}
\startdata
2.63 & H~{\sc i} (Br$\beta$) & --0.955 & 2.09(--12) $\pm$ 2.32(--13) & 4.1 $\pm$ 0.5 \\
2.86 & H~{\sc i} 5--11 & --0.962 & 6.12(--13) $\pm$ 8.41(--14) & 1.2 $\pm$ 0.2 \\
3.29 & PAH  & --0.971 & 8.81(--12) $\pm$ 2.55(--13) & 16.2 $\pm$ 1.0 \\
      & + H~{\sc i} 5--9             &         &                             &\\
3.41 & PAH & --0.973 & 1.47(--12) $\pm$ 2.98(--13) & 2.7 $\pm$ 0.6 \\
3.49 & PAH & --0.974 & 2.04(--12) $\pm$ 3.77(--13) & 3.7 $\pm$ 0.7 \\
3.56 & PAH & --0.975 & 6.92(--13) $\pm$ 3.47(--13) & 1.3 $\pm$ 0.6 \\
3.63 & H~{\sc i} 6--19 & --0.976 & 1.59(--12) $\pm$ 4.08(--13) & 2.9 $\pm$ 0.8 \\
      & + H~{\sc i} 6--20             &         &                             &\\
3.73 & H~{\sc i} 5--8 & --0.977 & 1.06(--12) $\pm$ 4.58(--13) & 1.9 $\pm$ 0.8 \\
3.82 & H~{\sc i} 6--16 & --0.978 & 1.19(--12) $\pm$ 8.38(--13) & 2.1 $\pm$ 1.5 \\
3.91 & H~{\sc i} 6--15 & --0.979 & 6.25(--13) $\pm$ 5.40(--13) & 1.1 $\pm$ 1.0 \\
4.05 & H~{\sc i} (Br$\alpha$)& --0.980 & 4.94(--12) $\pm$ 7.94(--14) & 8.9 $\pm$ 0.5 \\
      & + H~{\sc i} 6--14             &         &                             &\\
4.30 & He~{\sc i} 3--5                     & --0.982 & 6.97(--13) $\pm$ 6.52(--14) & 3.1$\pm$1.4 \\
4.38 & H~{\sc i} 6--12 & --0.982 & 5.72(--13) $\pm$ 7.07(--14) & 1.0 $\pm$ 0.1 \\
4.65 & H~{\sc i} (Pf$\beta$) & --0.984 & 8.99(--13) $\pm$ 1.17(--13) & 1.6 $\pm$ 0.2 \\
      & + H~{\sc i} 6--11             &         &                             &\\
8.5    & C$_{60}$$^{b}$ & --0.970 & 3.66(--12) $\pm$ 4.59(--13) & 6.8 $\pm$ 0.9 \\
8.6    & PAH$^{c}$            & --0.970 & 5.10(--12) $\pm$ 5.07(--13) & 9.4 $\pm$ 1.0 \\ 
8.99   & [Ar\,{\sc iii}] & --0.959 & 1.60(--12) $\pm$ 1.75(--13) & 3.1 $\pm$ 0.4 \\
12.37 & H~{\sc i} 6-7,8-11 & --0.980 & 9.68(--13) $\pm$ 4.24(--14) & 1.0 $\pm$ 0.1 \\
12.71 & He~{\sc i} & --0.982 & 4.85(--13) $\pm$ 5.35(--14) & 0.5 $\pm$ 0.1 \\
12.81 & [Ne~{\sc ii}] & --0.983 & 2.90(--11) $\pm$ 4.35(--13) & 30.9 $\pm$ 1.5 \\
17.3 &C$_{60}$$^{d}$ &--0.981 & 4.61(--12) $\pm$ 4.78(--13) & 4.9 $\pm$ 0.6\\
18.71 & [S~{\sc iii}] & --0.981 & 1.31(--12) $\pm$ 7.27(--14) & 1.4 $\pm$ 0.1 \\
18.9 & C$_{60}$$^{e}$  & --0.981 & 6.88(--12) $\pm$ 2.55(--13)& 7.4 $\pm$ 0.4 \\
33.16 & He~{\sc i} & --0.993 & 1.47(--12) $\pm$ 2.57(--13) & 1.5 $\pm$ 0.3 \\
35.83 & He~{\sc i} & --0.993 & 3.89(--12) $\pm$ 2.85(--13) & 4.0 $\pm$ 0.3
\enddata
\tablenotetext{a}{$\log$$F$(H$\beta$) = --11.63 erg s$^{-1}$ cm$^{-2}$.}
\tablenotetext{b}{The FWHM is 0.17 $\mu$m.}
\tablenotetext{c}{The sum of two Gaussian components representing the 
PAH 8.6 $\mu$m. See text in detail.}
\tablenotetext{d}{The FWHM is 0.51$\pm$0.04 $\mu$m.}
\tablenotetext{e}{The FWHM is 0.36$\pm$0.03 $\mu$m.}
\tablecomments{For interstellar reddening correction, we used $c$({\hb})=1.403$\pm$0.025
for the AKARI/IRC and 1.629$\pm$0.020 for the Spitzer/IRC spectra, respectively.}

\label{akari_sp}
\end{deluxetable}

\subsection{\emph{Spitzer}/IRS and VLT/VISIR archival data}
M1-11 was observed by \emph{Spitzer} on November 10, 2006 with 
the  9.9-19.6 $\mu$m (SH)
and 18.7-37.2 $\mu$m (LH) modes on the Infrared Spectrograph (IRS, Houck
et al. 2004), as part of 
program ID 30430 (PI: H.~Dinerstein; AORKEY: 19903232). 
We downloaded the archival spectral images, and after masking bad
pixels using IRSCLEAN, we extracted the one-dimensional spectra using
SPICE. The S/N is $>$30 for the dust continuum.

In the spectrum of Cohen et al. (1986), very weak and tentative features
are seen around 7.0 and 8.5 $\mu$m that may arise at least partially
from fullerene C$_{60}$ with possible blending from the PAH 8.5 $\mu$m
band. However, the quality of their data is insufficient to be confident of these features.
To check the presence of the C$_{60}$ 8.5 $\mu$m feature, 
we downloaded the 7.7-13.3 $\mu$m archival spectral data 
obtained using the VLT spectrometer and imager for the mid-infrared (VISIR) 
at ESO VLT UT3 (ID: 084.D-0868A; PI: E.Lagadec). We
reduced the raw data using ESO {\tt gasgano}. 
To compare the VISIR data with the \emph{Spitzer} spectra of other
C$_{60}$ PNe and also to combine it with the M1-11's \emph{Spitzer} 
spectrum, we degraded the 
original VISIR spectral resolving power of $\sim$400 down to 90 
by using a Gaussian convolution technique. The S/N of the convoluted VISIR spectrum is $>$70.

\begin{figure}
\centering
\epsscale{1.1}
\plotone{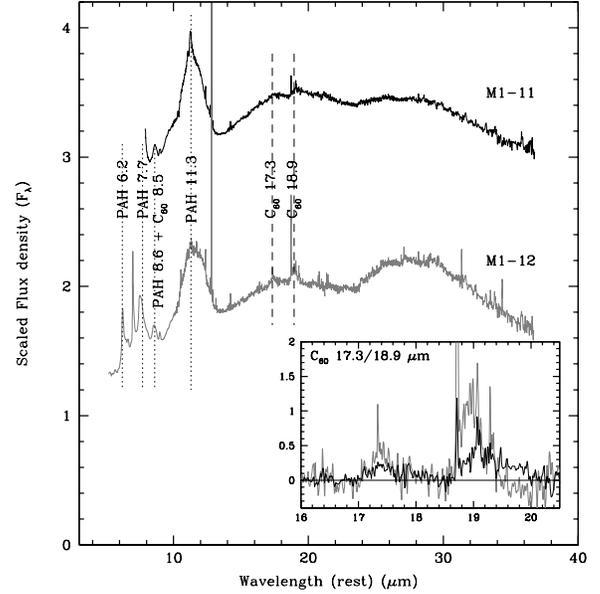}
\caption{The \emph{Spitzer} spectra of M1-11 (black line) and M1-12 (gray
 line). ({\it inner box}) The line-profiles of C$_{60}$ 17.33 and 18.94
 $\mu$m. The line-profiles of C$_{60}$ at 8.5 $\mu$m in M1-11 and M1-12
 are present in Fig.\ref{c60_spec}.}
\label{irs}
\end{figure}

The IRS and VISIR combined 7.7-37.2 $\mu$m spectrum which is shown in
Fig.~\ref{irs}, reveals the solid-state/molecular features and atomic lines on the dust continuum thermal emission.
The detected lines are listed in Table
\ref{akari_sp}. The emission around 8.5 $\mu$m is the complex 
of C$_{60}$ 8.5 $\mu$m and PAH 8.6 $\mu$m. We measured the 
flux density of C$_{60}$ 8.5 $\mu$m by using multiple Gaussian fitting
to separate its flux from the PAH 8.6 $\mu$m feature (See section 2.6.3). 
We performed the interstellar reddening correction in
a similar way as described in Section 2.5. By comparing the observed
intensity ratios H~{\sc i} $F$($n$=6--7)/$F$({\hb}) \&
$F$($n=$8--11)/$F$({\hb}) to the theoretical values of Storey \&
Hummer (1995) for the case B assumption in $T_{\rm e}$=10$^{4}$ K and
$n_{\rm e}$=10$^{4}$ cm$^{-3}$, we derived that $c$({\hb})=1.63$\pm$0.02.

\subsubsection{Broad spectral features at 10--13 and 16--22 $\mu$m}
M1-11 appears to have C-rich dust, as evidenced by the presence 
of a broad 10-13 $\mu$m feature, which is usually attributed to 
SiC in the literature. This feature (centered around 11.3 $\mu$m) 
is seen on top of a featureless continuum, presumably due to 
amorphous carbon (AC).
The PAH features around 10--11 $\mu$m are also visible in the
spectrum. A second broad feature is seen around approximately 16--22
$\mu$m, similar in appearance to broad features reported in several
PNe in the Magellanic Clouds (Stanghellini et al. 2007; Bernard-Salas
et al. 2009; Garc{\'i}a-Hern{\'a}ndez et al. 2011a; Garc{\'i}a-Hern{\'a}ndez et al. 2012)

While Stanghellini et al.~(2007) associate this emission feature with
carbon-rich dust, Bernard-Salas et al.~(2009) show that the 16--22
$\mu$m broad feature differs from the PAH plateau around 16--20 $\mu$m
(Van Kerckhoven et al. 2000) and the 21 $\mu$m feature sometimes seen
in carbon-rich post-AGB stars (e.g.~Volk et al.~2011) and PNe (Hony et
al.~2001), and is more similar to the 18 $\mu$m amorphous silicate
feature, thus assigning a oxygen-rich carrier for this feature in
objects.  Bernard-Salas et al.~(2009) imply that if the silicate
identification for this feature is correct, the Magellanic Cloud PNe
often show a dual dust chemistry, although earlier in the same study
they state that not a single source in their sample shows a mixed
chemistry, thus undermining the silicate identification of the 16--22
$\mu$m bump. Indeed, in a comprehensive study of a large sample of
Galactic and Magellanic Cloud PNe, Stanghellini et al. (2012) conclude
that none of the LMC PNe considered show a dual chemistry, lending
more credibility to the idea that the 16--22 $\mu$m feature is carried
by a carbon-based material.

Considering the gas chemistry adds to the confusion. Table
\ref{mcpnabund} shows the C and O gas-phase abundances derived from the
CELs for six of the PNe with 16--22 $\mu$m emission features discussed by
Bernard-Salas et al.~(2009) and Garc{\'i}a-Hern{\'a}ndez et
al. (2011a). Three of the six are O-rich in their gas-phase material,
while the remaining three MC PNe actually show a C-rich chemistry.
Since the C abundances of SMC1 and SMC6 have relatively large
uncertainty (0.25 dex), these two SMC PNe could be C-rich.
Even with this information, it remains unclear whether the
16--22 $\mu$m bump is due to an O-rich or a C-rich carrier. Thus we do
not assign any identification to this feature, and do not include
fitting the 16--22 $\mu$m bump in our analysis.

\begin{deluxetable}{ccccl}
\tablecolumns{5} 
\tablecaption{C and O abundances derived from
    CELs, for Magellanic Cloud PNe showing the
    16--22 $\mu$m feature. \label{mcpnabund}} 
\tablewidth{240pt}
\tablehead{ \colhead{Nebula}& \colhead{C$^{a}$}& \colhead{O$^{a}$}&
\colhead{C/O}& \colhead{Ref.}  } 
\startdata
  SMC1   &8.11&8.26&0.71&(1)\\
  SMC6   &8.03&8.22&0.65&(1)\\
  LMC8   &7.93&8.26&0.47&(2)\\
  LMC25   &8.29&8.17&1.32&(3),(4)\\
  LMC48   &8.40&8.24&1.45&(3),(4)\\
  LMC85 &8.74&8.40&2.19&(2)
\enddata
\tablerefs{(1) Idiart et al. (2007); (2) Dopita et al. (1997); (3)
 Stanghellini et al. (2005); (4) Leisy \& Dennefeld (2006)}
\tablenotetext{a}{The number density relative to the hydrogen is defined
 as $\log$~H=12.}
\end{deluxetable}

\subsubsection{30 $\mu$m broad feature}
The carrier of the 30 $\mu$m feature remains somewhat of a mystery.
While MgS has been proposed and often used as the carrier of this
feature (Hony et al.~2003), recent work has cast doubt on the
identification with MgS (see e.g., Zhang et al. 2009; 
Garc\'{i}a-Hern\'{a}ndez et al. 2010; Zhang \& Kwok 2011). Nevertheless,
we consider MgS as a possible dust component in M1-11 to explain the 30
$\mu$m feature in the spectrum.

Other potential carriers, in particular
hydrogenated amorphous carbon (HAC) (see Grishko et al. 2001; Hony et
al. 2003), should also be considered. In the SED model for proto-PN HD 56126,
Hony et al.~(2003) showed that broad emission features around 7-9 $\mu$m and
10-13 $\mu$m are partly due to HACs, making the identification of the
30 $\mu$m band with HACs (Grishko et al.~2001), a possibility.
We examined whether HAC can be the main contributor 
to the 10--13 and 30 $\mu$m broad features by 
applying a modified blackbody model to the IRS spectrum, including both 
MgS and HACs.

\begin{figure}
\centering
\epsscale{1.1}
\plotone{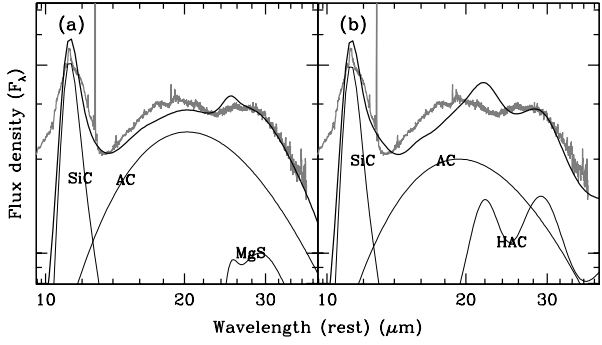}
\caption{The observed \emph{Spitzer}/IRS and VLT/VISIR spectrum of M1-11 (gray
  lines) and the predicted SED from modified blackbody fitting (thick
  lines). Each component is indicated by a thin line. AC stands
 for amorphous carbon.
\label{bb_fig} }
\end{figure}

The observed flux density $F_{\lambda}$ due to thermal emission 
from dust grains is given by
\begin{equation}
F_{\lambda} = \sum_{i}\left(\frac{4}{3}\pi a_{i}\rho_{i} D^{2}\right)^{-1}
m_{d,i}Q_{\lambda,i}\pi B_{\lambda}(T_{d,i}),
\label{dust}
\end{equation}
\noindent
where $a_{i}$ is the grain radius of component $i$, $\rho_{i}$ is the
dust density, $m_{d,i}$ is the dust mass, $Q_{\lambda,i}$ is the absorption
efficiency, and $B_{\lambda}$($T_{d,i}$) is the Planck function for a
dust temperature $T_{d,i}$.  We adopt a distance $D = 2.1$
kpc (Tajitsu \& Tamura 1998). In Table \ref{optdata} we list the optical 
constants that we use for each of the dust species considered. For MgS, we 
used the optical constants of nearly pure MgS, e.g.~Mg$_{0.9}$Fe$_{0.1}$S, 
from Begemann  et al. (1994).
For the HACs, we
adopt the $Q_{\lambda}$ of HAC in the case of H/(H+C)=0.3 calculated
by Hony et al. (2003). 
In two models, we consider different compositions, consisting of 
combinations of PAHs, amorphous carbon (AC),
SiC, MgS, and HAC (Fig.~\ref{bb_fig}). We considered 9.9-37.2 $\mu$m
\emph{Spitzer} spectrum and \emph{AKARI} FIS 65/90 $\mu$m photometry
data, except for the 16-22 $\mu$m broad band. We assume spherically 
shaped grains with a radius of $a$=0.5 $\mu$m for AC, SiC, and HACs, and 
we excluded PAHs.  To reproduce the broad 30 $\mu$m emission
using MgS, as discussed in Hony et al.~(2003), we considered a continuous
distribution of ellipsoids (CDE, e.g., Bohren \& Huffman 1983; Fabian et
al. 2001; Min et al. 2003) and calculated $Q_{\lambda}$ of CDE MgS
using Eq.~18 given by Min et al.~(2003). To simplify, we assume that
the value of each ellipsoid MgS grain is $\simeq$4$\pi$$a^{3}$/3, where
$a$ is 0.5 $\mu$m.

The results of the modified blackbody fitting is shown in
Fig.~\ref{bb_fig}, and the derived $T_{d,i}$, $m_{d,i}$, and mass
fraction for each dust component are summarized in Table \ref{bb_res}.
Model 1 with AC, SiC, and MgS can explain the observed spectrum
reasonably well,
while the model 2 with HAC instead of MgS predicts an unseen broad 
emission feature around 20 $\mu$m. 
However, we should keep in mind 
that it is extremely difficult to characterize carbon compounds with 
a mixed aromatic and aliphatic content - such as HACs - in the
laboratory because these optical constants are strongly variable for different chemical 
and physical conditions. Jones (2012) presents HAC theoretical 
models 
 that show the extreme variability 
of HAC spectra depending on parameters such as hydrogen-content, 
grain size, etc. In any case, the 20 $\mu$m feature is much weaker than
the 30 $\mu$m feature (see Ghrisko et al. 2001) and one could detect the
30 $\mu$m features while not detecting the 20 $\mu$m feature. 
At present, therefore, we do not completely rule out that HACs as 
a carrier of the 30 $\mu$m broad feature.

In this paper, therefore, we assume that SiC and MgS are the main
contributors to the 10--13 and 30 $\mu$m broad features, respectively. 
In the remainder of our analysis, we will only consider PAHs, 
SiC, AC, and MgS to model the dust
emission in the SED.

\begin{deluxetable}{@{}ll@{}}
\tablecolumns{2}
\tablecaption{Adopted optical constants for each model dust component.\label{optdata}}
\tablewidth{240pt}
\tablehead{
\colhead{Dust species}&
\colhead{Data source}
}
\startdata
SiC&Pegourie (1988)\\
amorphous carbon (AC) &Rouleau $\&$ Martin (1991)\\
MgS&Begemann  et al. (1994)\\
HAC&Hony et al. (2003)
\enddata
\end{deluxetable}

\begin{deluxetable}{@{}cccc@{}}
\tablecolumns{4}
\centering
\tablecaption{Results from the modified blackbody fitting to the \emph{Spitzer}/IRS spectrum.\label{bb_res}}
\tablewidth{240pt}
\tablehead{
\colhead{Models}&
\colhead{Dust}&
\colhead{$T_{d}$}&
\colhead{$m_{d}$}\\
\colhead{}&
\colhead{Comp.}&
\colhead{(K)}&
\colhead{($M_{\odot}$)}
}
\startdata
model 1            &SiC  & 160 &2.85(--5)   \\
(Fig.\ref{bb_fig}a)&AC   & 120 &2.97(--4)   \\
                   &MgS  & 220 &3.17(--6)   \\
\hline
model 2            &SiC & 170 &1.75(--5)   \\
(Fig.\ref{bb_fig}b)&AC & 120  &1.77(--4)   \\
                   &HAC  & 80  &0.11        
\enddata
\end{deluxetable}

\subsubsection{8.5, 17.3, and 18.9 $\mu$m emission due to C$_{60}$ fullerenes} 
In the VISIR and Spitzer combined spectrum of M1-11, we see infrared
features at 8.5 (although blended with the PAH 8.6 $\mu$m, see below), 17.3, 
and 18.9 $\mu$m, most likely due to the fullerene C$_{60}$.

\begin{figure}
\centering
\epsscale{1.1}
\plotone{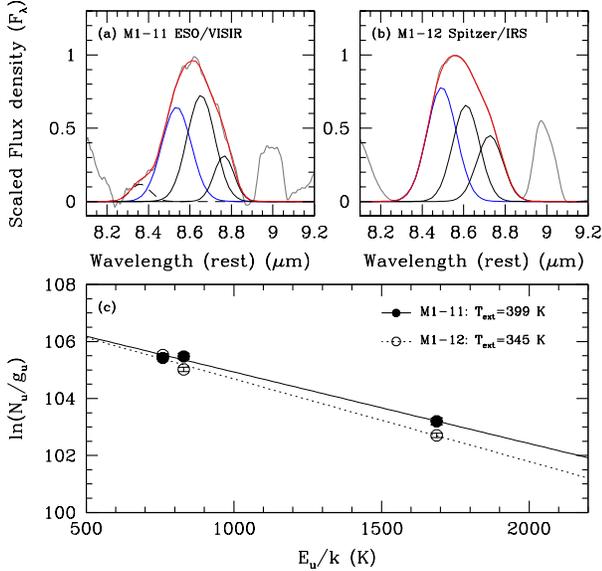}
\caption{({\it panels (a) and (b)}) The line-profiles of C$_{60}$ at 8.5
 $\mu$m in M1-11 and M1-12 (blue lines). The local dust continua are subtracted. The gray lines are the observations. We fit  the broad line at 8.6
 $\mu$m by three or four Gaussian. The deconvolved  profiles are
 indicated by the black and blue lines. The red lines are the sum of
 these components. The emission-line around 9 $\mu$m is [Ar\,{\sc iii}]
 8.99 $\mu$m. ({\it panel (c)}) The excitation diagram for C$_{60}$ in
 M1-11 and M1-12. The filled and open circles are the observed data in
 M1-11 and M1-12, respectively. The lines indicated represent the best
 fit to the data. See text in detail.
\label{c60_spec} }
\end{figure}

In our own Milky Way Galaxy, these C$_{60}$ infrared features were
recently detected in five PNe, including Tc1, M1-12, M1-20, 
K3-54, and M1-60 
(Cami et al. 2010; Garc\'{i}a-Hern\'{a}ndez et al. 2010; Garc\'{i}a-Hern\'{a}ndez et al. 2012), 
a C-rich PPN (IRAS01005+7910; Zhang \& Kwok 2011), and two O-rich 
post-AGB stars (IRAS06338+5333 and HD52961; Gielen et al. 2011). In
Fig.\ref{irs}, we show the IRS spectrum of M1-12 in comparison to
M1-11. The C$_{60}$ 17.3 and 18.9 $\mu$m features seem to be present 
in M1-11 although they are much weaker than those in M1-12. The spectrum 
of M1-12 shows spectral features due to PAHs, SiC, AC, MgS, and a very
weak 16--22 $\mu$m feature, which resembles that of M1-11. The 16--22 
$\mu$m feature is not seen in Tc1, M1-20, K3-54, and M1-60. Some other PNe 
with fullerenes in the Magellanic Clouds exhibit the 16--22 um feature 
(Garc\'{i}a-Hern\'{a}ndez  et al. 2011a; Garc\'{i}a-Hern\'{a}ndez  et al. 2012).
In the inset of Fig.\ref{irs}, the $\sim$16-20 $\mu$m spectra of M1-11 
and M1-12 with a local dust continuum subtracted, are shown. 
The wavelength positions
of the intensity peak and the line widths of C$_{60}$ 17.3 and 18.9
$\mu$m lines in M1-11 are almost coincident with those in M1-12.
In addition, the complex line around 8.5-8.6 $\mu$m - 
which turns out to be a blend of PAH 8.6 $\mu$m and C$_{60}$ 8.5 $\mu$m 
- in M1-11 is very similar to that in M1-12, as shown in
Fig.~\ref{c60_spec}. These comparison with M1-12 supports 
the identification of this feature as a C$_{60}$+PAH blend.

To quantify the excitation temperature and the total number of C$_{60}$, 
we need to separate flux of C$_{60}$ 8.5 
$\mu$m from the PAH 8.6 $\mu$m band. Cami et al. (2010) reported that 
the FWHM of C$_{60}$ 8.5 $\mu$m in the PN Tc1 is 0.15 $\mu$m 
by Gaussian fitting. Tc1 shows strong C$_{60}$ lines but very weak PAH bands. 
Assuming that the FWHM of C$_{60}$ 8.5 $\mu$m is
$\sim$0.15 $\mu$m in M1-11 and M1-12, 
we fit the broad line at 8.6 $\mu$m with multiple 
Gaussians. The line at 8.6 $\mu$m in M1-11 can be represented by 
three plus one Gaussian components, as shown in Fig.\ref{c60_spec}a. 
We assume that the profile of the PAH 8.6$\mu$m line 
is represented by the sum of two Gaussians at the peak wavelengths 
of $\sim$8.7 and $\sim$8.8 $\mu$m. The FWHM of the PAH 8.6
$\mu$m represented by the sum of these two Gaussian is 0.2 $\mu$m, which
 is consistent with NGC7027 (0.23 $\mu$m in ISO/SWS spectrum
 shown in Fig.~\ref{akari}). The FWHM of the 
C$_{60}$ 8.5 $\mu$m line indicated by the blue line is 0.17 $\mu$m. 
The FWHM of the C$_{60}$ and PAH complex is 0.3 $\mu$m. The resultant
Gaussian fitting for M1-12 is presented in Fig.\ref{c60_spec}b. The PAH
8.6 $\mu$m is represented by the sum of two Gaussians at $\sim$8.6 and
$\sim$8.7 $\mu$m. The FWHMs of the C$_{60}$ and the PAHs for M1-12 are as same
as those of M1-11. The measured fluxes of the C$_{60}$ 8.5 $\mu$m and
PAH 8.6 $\mu$m lines in M1-11 are listed in Table \ref{akari_sp}.

The excitation temperature and the number of C$_{60}$ were derived by 
creating a vibration excitation diagram as shown in Fig.\ref{c60_spec}c. 
We followed the method of Cami et al. (2010). 
$N_{u}$ is the number of C$_{60}$ molecules in the upper
vibrational levels. $N_{u}$ is written by 

\begin{equation}
N_{u} = \frac{4 \pi I(C_{60}) D^{2}}{A} \frac{\lambda}{hc},
\label{c60_eq1}
\end{equation}
\noindent
where $I$(C$_{60}$) is the fluxes of C$_{60}$ lines in erg
s$^{-1}$ cm$^{-2}$, $D$ is the distance to M1-11 (2.1 kpc; Tajitsu \&
Tamura 1998), $A$ are the transition probabilities (4.2,
1.1, 1.9 s$^{-1}$ for C$_{60}$ 8.5, 17.3, and 18.9 $\mu$m,
respectively, from Garc\'{i}a-Hern\'{a}ndez et al. 2011b), 
$h$ is Planck's constant, and $c$ is 
the speed of light. The vibrational degeneracy is given by $g_{u}$. In thermal
equilibrium, the Boltzmann equation relates the $N_{u}$ to the 
excitation temperature $T_{ext}$:
\begin{equation}
N_{u} \propto g_{u}\exp(-E{u}/kT_{ext}).
\label{c60_eq2}
\end{equation}
$E{u}$ and $k$ are the energy of the excited level and the
Boltzmann constant, respectively. We confirmed that 
our measured excitation temperature, $T_{ext}$ of 338$\pm$9 K in Tc 1 using
Eqs. (\ref{c60_eq1}) and (\ref{c60_eq2}) is consistent with Cami et al. (2010; 332
K). Accordingly, we obtained $T_{ext}$ of 399$\pm$36 K and the
$N$(C$_{60}$) of 4.57$\pm$1.23(+46). 
The total mass of C$_{60}$ $m_{\rm C60}$ is 2.75(--8) $M_{\odot}$.

Our estimated $T_{ext}$, $N$(C$_{60}$), and $m_{\rm C60}$ in M1-12 are
345$\pm$35 K, 5.30$\pm$1.23(+46), and 3.18(--8) $M_{\odot}$,
respectively, adopting $D$ of 3.9 kpc (Tajitsu \& Tamura 1998). Garc\'{i}a-Hernandez 
et al. (2010) measured the $T_{ext}$ in M1-12 of  546 K.
Their measured FWHM of C$_{60}$ 8.5 $\mu$m is 0.237 $\mu$m. The
$T_{ext}$ discrepancy between theirs is due to the differences in the 
measured line flux of this line.

\section{Results}

\subsection{CELs plasma diagnostic}
We determined the electron temperatures {\te} and densities {\Ne} using
11 diagnostic CELs, and listed the results in Table \ref{diano_table}.

For {\nii} $\lambda$5755 and {\oii}$\lambda\lambda$7320/30, we
subtracted the recombination contamination from both lines
using 
\begin{equation}
\label{rni}
\frac{I_{R}(\rm [N\,{\sc II}]\lambda5755)}{I(\rm H\beta)} =
3.19\left(\frac{T_{\rm e}}{10^4}\right)^{0.33}\times\frac{\rm N^{2+}}{\rm H^{+}},
\end{equation}
and
\begin{equation}
\label{roii}
\frac{I_{R}(\rm [O\,{\sc II}]\lambda\lambda7320/30)}{I(\rm H\beta)} =
9.36\left(\frac{T_{\rm e}}{10^4}\right)^{0.44}\times\frac{\rm O^{2+}}{\rm H^{+}}.
\end{equation}
given by Liu et al. (2000).  Adopting N$^{2+}$ and O$^{2+}$ ionic
abundances derived from N~{\sc ii} and O~{\sc ii} lines (see Section
\ref{orlabun}), we derived that $I_{R}$({\rm {\nii}} $\lambda$5755) =
0.12 and $I_{R}$({\rm {\oii}} $\lambda\lambda$7320/30) = 2.11.  The
recombination contamination is $\sim$2 $\%$ in {\nii} $\lambda$5755
and $\sim$8 $\%$ in {\oii}, respectively.

\begin{figure}
\centering
\epsscale{1.1}
\plotone{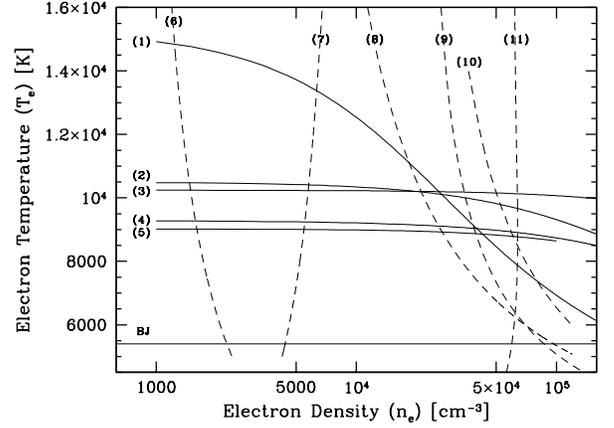}
\caption{{\Ne}-{\te} diagram. Each curve is labeled with an ID
 number given in Table \ref{diano_table}.  The solid lines indicate
 diagnostic lines of the {\te}. The broken lines indicate diagnostic
 lines of the {\Ne}. }
\label{diagno_figure}
\end{figure}

The resulting {\Ne}--{\te} diagnostic diagram is shown in
Fig.~\ref{diagno_figure}.  The solid lines indicate diagnostic lines
for the electron temperatures, while the broken lines are electron
density diagnostics.  Since the gas in M1-11 has  much higher density than
the critical densities  of the density sensitive lines {\oii}
$\lambda\lambda$3726/29 and {\sii} $\lambda\lambda$6717/31, the ratios
of {\oii} $I$($\lambda\lambda$3726/29)/$I$($\lambda\lambda$7320/30)
and {\sii} $I$($\lambda\lambda$6717/31)/$I$($\lambda\lambda$4069/76)
are density-sensitive rather than temperature-sensitive.  The critical
densities at {\te}=10\,000 K for {\oii} $\lambda\lambda$3726/29 are
$\sim$4500 and $\sim$980 cm$^{-3}$, respectively, and those of {\sii}
$\lambda\lambda$6717/31 are $\sim$1400 and $\sim$3600 cm$^{-3}$,
respectively.  The densities derived from the {\oii}
$\lambda\lambda$3726/29 and {\sii} $\lambda\lambda$6717/31 ratios
might be the value for thin shell region, while those from the {\oii}
$I$($\lambda\lambda$3726/29)/$I$($\lambda\lambda$7320/30) and {\sii}
$I$($\lambda\lambda$6717/31)/$I$($\lambda\lambda$4069/76) would be the
value for the bright rim.  To determine {\Ne}, we adopted
{\te}=10\,000 K for all density-diagnostic lines. {\te}({\nii}) was
calculated using {\Ne}=36\,490 cm$^{-3}$, which is the average between
{\Ne}({\oii}) derived from
$I$($\lambda\lambda$3726/29)/$I$($\lambda\lambda$7320/30)
({\Ne}({\oii}$_{n/a}$) hereafter) and {\Ne}({\siii}).  For
{\te}({\siii}), {\te}({\oiii}), and {\te}({\ariii}), we adopted
{\Ne}({\siii}).  For {\te}({\oi}), we adopted {\Ne}({\Ni}). Our
estimates for {\te}({\nii}), {\te}({\oiii}), and {\te}({\siii}) are in
excellent agreement with those by Henry et al. (2010), who estimated
10\,720, 9996, and 9100 K, respectively. The estimates for {\Ne} and
{\te} are summarized in Table \ref{diano_table}.

To calculate the CEL ionic abundances, we adopt a three zone model
based on {\Ne}-{\te} diagram. The {\te} and {\Ne} combinations
for each ion are listed in Table \ref{temp_ne_cles}. For the N$^{0}$
and O$^{0}$ abundances, we adopted {\te}({\oi}) and {\Ne}({\Ni}) (zone
1). The averaged {\Ne} is from {\Ne}({\oii}$_{n/a}$) and
{\Ne}({\siii}); and {\te}({\nii}) is for ions with 0$<$IP$\lesssim$13.6 eV
(zone 2).  {\Ne}({\siii}) and the averaged temperature among
{\te}({\oiii}), {\te}({\siii}), and {\te}({\ariii}) is for ions with
I.P.=14.4--35.5 eV (zone 3).

\begin{deluxetable}{@{}clcc@{}}
\tablecolumns{4}
\tabletypesize{\small}
\tablecaption{Adopting {\Ne} and {\te} for the CEL ionic abundance calculations.\label{temp_ne_cles}}
\tablewidth{240pt}
\tablehead{
\colhead{Zone}&
\colhead{Ions}&
\colhead{{\Ne}(cm$^{-3}$)}&
\colhead{{\te}(K)}
}
\startdata
1 &N$^{0}$, O$^{0}$ &1380   &9240\\
2 &N$^{+}$, O$^{+}$, P$^{+}$, S$^{+}$, Cl$^{+}$       &36\,490   &8410\\
3 &O$^{2+}$, Ne$^{+}$, S$^{2+}$, Cl$^{2+}$,           &51\,120   &9540\\
  &Ar$^{2+}$, Fe$^{2+}$, Kr$^{2+}$                               &          &
\enddata
\end{deluxetable}

\subsection{RL plasma diagnostics}
We calculate the He, C, N, and O abundances using RLs of
these elements by adopting the
{\te} derived from the Balmer discontinuity
together with {\hei} line ratios and the {\Ne} from the Balmer decrements,
listed in Table \ref{diano_table}.
The Balmer discontinuity temperature {\te}(BJ) was determined using
the method described by Liu et al. (2001), which we used to obtain the C$^{2+}$,
N$^{2+}$, and O$^{2+}$ abundances from recombination lines.

The {\hei} electron temperatures {\te}({\hei}) were derived from the
ratios of He\,{\sc i} $I$($\lambda$7281)/$I$($\lambda$6678),
$I$($\lambda$7281)/$I$($\lambda$5876), and
$I$($\lambda$6678)/$I$($\lambda$5876) assuming a constant electron
density of 10$^{6}$ cm$^{-3}$, estimated from the Balmer decrements
(see below). We adopted the emissivities of {\hei} by Benjamin et al.
(1999). The {\te}({\hei}) derived from three different line ratio
combinations is 3980-6980 K. We adopted {\te}({\hei}) derived from
He\,{\sc i} $I$($\lambda$7281)/$I$($\lambda$6678) for the He$^{+}$
abundance calculations.  The reason why this {\te}({\hei}) is the most
reliable was discussed in Otsuka et al. (2010).

\begin{figure}
\centering
\epsscale{1.1}
\plotone{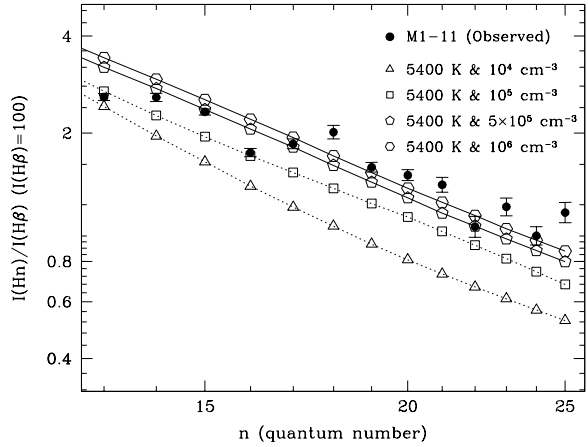}
\caption{
Plot of the intensity ratio of the higher order Balmer lines
to {\hb} (Case B assumption) with the theoretical intensity
 ratios in {\te}=5400 K and different {\Ne}.
\label{bal}}
\end{figure}

The intensity ratios of the high-order Balmer lines Hn (n: the
principal quantum number of the upper level) to a lower Balmer line,
e.g., {\hb}, are also sensitive to the electron density.  In
Fig.~\ref{bal}, we plot the ratios of higher-order Balmer lines to
{\hb} compared to the theoretical values by Storey \& Hummer (1995)
for {\te}(BJ) and {\Ne} of 10$^{5}$, 5$\times$10$^{5}$, and 10$^{6}$
cm$^{-3}$. The electron density in the recombination line emitting
regions seems to be $>$10$^{5}$ cm.
Care when dealing with this
value is necessary, because it apparently has large scatter.

\begin{deluxetable*}{lllrl}
\tablecolumns{5}
\tablecaption{Plasma diagnostic results.\label{diano_table}}
\tablewidth{480pt}
\tablehead{
\colhead{}   &
\colhead{ID}&
\colhead{Diagnostic}&
\colhead{Value}&
\colhead{Result}
}
\startdata
$T_{\rm e}$ &(1)&[N\,{\sc ii}] ($\lambda$6548/83)/($\lambda$5755)&42.45$\pm$1.14$^{a}$&8410$\pm$90\\

(K)
&(2)&[O\,{\sc iii}] ($\lambda$4959/5007)/($\lambda$4363)&182.94$\pm$20.32&9740$\pm$330\\
 &(3)&[Ar\,{\sc iii}]($\lambda$7135)/($\lambda$5192)&132.44$\pm$26.92&10\,060$\pm$780\\
&(4)&[O\,{\sc i}] ($\lambda$6300/63)/($\lambda$5577)&74.62$\pm$14.31&9240$\pm$640\\
&(5)&[S\,{\sc iii}] ($\lambda$9069)/($\lambda$6313)&9.96$\pm$2.63&8830$\pm$980\\
\cline{2-5}
 & &He\,{\sc i} ($\lambda$7281)/($\lambda$6678)&0.20$\pm$0.01&6890$\pm$330\\
 & &He\,{\sc i} ($\lambda$7281)/($\lambda$5876)&0.052$\pm$0.002&5920$\pm$230\\
 & &He\,{\sc i} ($\lambda$6678)/($\lambda$5876)&0.26$\pm$0.01&3980$\pm$160\\
 & &Average &&5600\\
\cline{2-5}
 &BJ &Balmer Jump&&5400$\pm$1300\\
\hline
$n_{\rm e}$     &(6)&[N\,{\sc i}] ($\lambda$5198)/($\lambda$5200)&1.63$\pm$0.13&1380$\pm$350\\
(cm$^{-3}$)     &(7)&[O\,{\sc ii}] ($\lambda$3726)/($\lambda$3729)&2.75$\pm$0.05&5750$\pm$380\\
                &(8)&[O\,{\sc ii}] ($\lambda$3626/29)/($\lambda$7320/30)&3.11$\pm$0.10$^{b}$&21\,860$\pm$780\\
 &(9)&[S\,{\sc ii}]($\lambda$6716/31)/($\lambda$4069/76)&0.63$\pm$0.02&35\,350$\pm$1400\\
                &(10)&[S\,{\sc iii}] ($\lambda$18.71$\mu$m)/($\lambda$9069)&0.44$\pm$0.11&51\,120$\pm$15\,550\\
                &(11)&[Cl\,{\sc iii}] ($\lambda$5517)/($\lambda$5537)&0.40$\pm$0.10&22\,110-63\,730\\
\cline{2-5}
&&Balmer decrements&&10$^{5}$--10$^{6}$
\enddata
\tablenotetext{a}{Corrected recombination contribution for $[$N\,{\sc ii}$]$ $\lambda$5755.}
\tablenotetext{b}{Corrected recombination contribution for $[$O\,{\sc ii}$]$ $\lambda\lambda$7320/30.}
\end{deluxetable*}

\subsection{CEL ionic abundances}
The derived ionic abundances are listed in Table \ref{cel_abund}.  In
the last line of the transition series for each ion, we present the
adopted ionic abundance in bold face. The adopted values represent the
line intensity weighted mean in case two or more lines are detected.
As references, the results by Sterling \& Dinerstein (2008, for Kr)
and Henry et al. (2010, for the others) are also listed in the last
column. This is the first time the Ne$^{+}$, P$^{+}$, and Fe$^{2+}$
abundances are derived for M1-11. In total, fifteen ionic abundances
are determined by solving for a $>$5 level atomic model, with the
exception of Ne$^{+}$, for which the abundance was calculated using a
two level energy model. We adopted the same collisional strengths and
transition probabilities used in Otsuka et al. (2010, 2011) except for
Cl$^{+}$, for which we adopted the transition probabilities from the
CHIANTI atomic
database\footnote[9]{http://www.ukssdc.ac.uk/solar/chianti/}, the
collisional impacts of Tayal (2004b), and the level energy listed in
Atomic Line List v2.05b12.  We subtracted the recombination
contamination in the {\nii}$\lambda$5755 and
{\oii}$\lambda\lambda$7320/30 lines to derive the N$^{+}$ and O$^{+}$
abundances.

In general, our derived abundances are comparable to the values of
Henry et al.~(2010) and Sterling \& Dinerstein (2008).  The
discrepancies between our N$^{+}$, O$^{+}$, S$^{+}$ abundances and the
results by Henry et al.~(2010) are mainly due to adopted electron
temperature; Henry et al.~adopted 10\,200 K. Note that they employed
three different {\te}s and a constant {\Ne}=20\,000 cm$^{-3}$ in their
models.  With their {\te}s, we would find N$^{+}$, O$^{+}$, and
S$^{+}$ abundances comparable to their results. Lastly, a minor
discrepancy for the N$^{+}$ and O$^{+}$ abundances must be due to the
recombination contamination in the {\nii}$\lambda$5755 and
{\oii}$\lambda\lambda$7320/30 lines, respectively.

\begin{deluxetable}{@{}c@{\hspace{2pt}}c@{\hspace{2pt}}c@{\hspace{2pt}}c|l@{}}
\tablecolumns{5}
\tablecaption{Ionic abundances derived from CELs.\label{cel_abund}}
\tablewidth{240pt}
\tablehead{
\colhead{X$^{m+}$}&
\colhead{$\lambda_{\rm lab}$}&
\colhead{$I$($\lambda$)}&
\colhead{X$^{m+}$/H$^{+}$}&
\colhead{Others}\\
\colhead{}&
\colhead{({\AA} or $\mu$m)}&
\colhead{[$I$({\hb})=100]}&
\colhead{}&
\colhead{}
}
\startdata
N$^{0}$ & 5197.90 & 2.01(--1) $\pm$ 9.10(--3) & 3.25(--7) $\pm$ 7.79(--8)&\nodata \\
  & 5200.26 & 1.23(--1) $\pm$ 7.78(--3) & 3.21(--7) $\pm$ 7.54(--8)&\nodata  \\
   &  &  & {\bf 3.24(--7) $\pm$ 7.69(--8)} &\nodata\\
N$^{+}$ & 5754.64 & 5.87(0) $\pm$ 1.08(--1) & 1.14(--4) $\pm$ 6.93(--6)&4.32(--5) \\
 & 6527.24 & 3.02(--2) $\pm$ 4.27(--3) & 7.25(--5) $\pm$ 1.05(--5) &\nodata\\
  & 6548.04 & 5.89(+1) $\pm$ 9.89(--1) & 8.41(--5) $\pm$ 3.08(--6) &4.13(--5)\\
  & 6583.46 & 1.90(+2) $\pm$ 4.77(0) & 9.18(--5) $\pm$ 3.77(--6) &4.32(--5)\\
 &  &    &{\bf  9.05(--5) $\pm$ 3.68(--6)} &4.28(--5)\\
N$^{2+}$ & 1750 & 1.48(0) $\pm$ 5.08(--1) &{\bf 3.10(--5) $\pm$ 1.85(--5)}&\nodata \\
O$^{0}$ & 5577.34 & 3.95(--2) $\pm$ 7.56(--3) & 5.87(--6) $\pm$ 2.39(--6)&\nodata \\
  & 6300.30 & 2.20(0) $\pm$ 4.08(--2) & 5.79(--6) $\pm$ 1.34(--6)&3.26(--6) \\
  & 6363.78 & 7.48(--1) $\pm$ 2.13(--2) & 6.15(--6) $\pm$ 1.43(--6)&3.57(--6) \\
 &  &    &{\bf  5.88(--6) $\pm$ 1.38(--6)}&3.34(--6) \\
O$^{+}$ & 3726.03 & 5.54(+1) $\pm$ 5.26(--1) & 3.87(--4) $\pm$ 2.12(--5)&6.27(--5)$^{a}$ \\
  & 3728.81 & 1.97(+1) $\pm$ 3.02(--1) & 5.02(--4) $\pm$ 2.83(--5)&\nodata \\
 & 7320/30 & 2.37(+1) $\pm$ 6.92(--1) & 3.40(--4) $\pm$ 2.67(--5)&7.15(--5) \\
 &  &    &{\bf  3.99(--4) $\pm$ 2.39(--5)} & 7.04(--5)\\
O$^{2+}$ & 4361.21 & 1.37(--1) $\pm$ 1.52(--2) & 8.83(--6) $\pm$ 3.50(--6)&5.77(--6) \\
  & 4958.91 & 6.44(0) $\pm$ 3.09(--2) & 8.27(--6) $\pm$ 1.81(--6) &5.44(--6)\\
  & 5006.84 & 1.85(+1) $\pm$ 8.17(--2) & 8.25(--6) $\pm$ 1.80(--6) &5.77(--6)\\
 &  &    &{\bf  8.26(--6) $\pm$ 1.81(--6)} &5.69(--6)\\
Ne$^{+}$ & 12.81 & 3.09(+1) $\pm$ 1.45(0) &{\bf 4.76(--5) $\pm$ 2.98(--6)}&\nodata \\
P$^{+}$ & 1.15 & 1.22(--1) $\pm$ 1.82(--2) & 2.19(--8) $\pm$ 3.30(--9) &\nodata\\
 & 1.19 & 3.41(--1) $\pm$ 1.54(--2) & 2.32(--8) $\pm$ 1.14(--9) &\nodata\\
 &  &    &{\bf  2.29(--8) $\pm$ 1.71(--9)} &\nodata\\
S$^{+}$ & 4068.60 & 1.79(0) $\pm$ 4.65(--2) & 4.49(--7) $\pm$ 2.19(--8) &\nodata\\
  & 4076.35 & 6.34(--1) $\pm$ 5.55(--2) & 4.74(--7) $\pm$ 4.58(--8) &\nodata\\
  & 6716.44 & 4.42(--1) $\pm$ 1.61(--2) & 3.63(--7) $\pm$ 1.72(--8) &1.30(--7)\\
  & 6730.81 & 1.08(0) $\pm$ 2.68(--2) & 4.15(--7) $\pm$ 1.62(--8) &1.46(--7)\\
 &  &    &{\bf  4.34(--7) $\pm$ 2.37(--9)} &1.41(--7)\\
S$^{2+}$ & 6312.10 & 3.23(--1) $\pm$ 1.50(--2) & 8.45(--7) $\pm$ 2.23(--7)&7.67(--7) \\
  & 9068.60 & 3.22(0) $\pm$ 7.49(--1) & 1.02(--6) $\pm$ 2.73(--7) &\nodata\\
  & 18.71 & 1.40(0) $\pm$ 1.00(--1) & 9.59(--7) $\pm$ 9.04(--8) &\nodata\\
 &  &  &   {\bf 9.91(--7) $\pm$ 9.08(--8)} &7.68(--7)\\
Cl$^{+}$&8578.69&1.19(--1) $\pm$ 2.62(--2)&8.11(--9) $\pm$	1.80(--9)&1.29(--8)\\
	&9123.60&7.76(--2) $\pm$ 2.74(--2)&2.03(--8) $\pm$	7.18(--9)&\nodata\\
	&	&         &{\bf 1.29(--8) $\pm$ 3.92(--9)}&1.29(--8)\\
Cl$^{2+}$ & 5517.66 & 2.88(--2) $\pm$ 5.72(--3) & 1.51(--8)  $\pm$ 4.15(--9) &\nodata\\
  & 5537.60 & 7.22(--2) $\pm$ 1.15(--2) & 1.55(--8) $\pm$ 3.85(--9) &\nodata\\
 &  &  &   {\bf 1.54(--8) $\pm$ 3.94(--9)} &\nodata\\
Ar$^{2+}$ & 5191.82 & 1.86(--2) $\pm$ 3.76(--3) & 2.91(--7) $\pm$ 1.07(--7) &\nodata\\
  & 7135.80 & 2.46(0) $\pm$ 4.18(--2) & 2.20(--7) $\pm$ 2.51(--7) &3.90(--7)\\
   & 8.99 & 3.06(0) $\pm$ 3.76(--1) & 4.38(--7) $\pm$ 5.85(--8) &\nodata\\
 &  &  &   {\bf 2.41(--7) $\pm$ 3.95(--8)} &1.86(--7)\\
Fe$^{2+}$ & 4701.53 & 6.89(--2) $\pm$ 8.46(--3) & 9.31(--8) $\pm$ 2.28(--8)&\nodata \\
  & 4754.69 & 3.86(--2) $\pm$ 8.65(--3) & 1.26(--7) $\pm$ 3.90(--8)&\nodata \\
 & 4769.43 & 5.32(--2) $\pm$ 6.12(--3) & 2.08(--7) $\pm$ 5.03(--8)&\nodata \\
  & 4881.00 & 4.93(--2) $\pm$ 5.38(--3) & 3.74(--8) $\pm$ 9.27(--9)&\nodata \\
 & 4934.08 & 1.63(--2) $\pm$ 3.54(--3) & 1.49(--7) $\pm$ 4.42(--8)&\nodata \\
 & 5270.40 & 7.06(--2) $\pm$ 3.70(--3) & 8.20(--8) $\pm$ 1.63(--8)&\nodata \\
 &  &    & {\bf 1.09(--7) $\pm$ 2.72(--8)} &\nodata\\
Kr$^{2+}$ & 2.19 & 8.84(--2) $\pm$ 7.28(--3) & {\bf 3.52(--9) $\pm$ 3.87(--10)}&2.99(--9)
\enddata
\tablenotetext{a}{Based on {\oii}$\lambda\lambda$3726/29 complex.}
\tablecomments{(in fifth column) All values except for Kr$^{2+}$
are from Henry et al. (2010). The Kr$^{2+}$ abundance is from Sterling
\& Dinerstein (2008).}
\end{deluxetable}

\subsection{RL ionic abundances \label{orlabun}}
Our values for the ionic abundances derived from RLs are listed in
Table \ref{orl_abun}.  In general, the Case B assumption applies to
lines from levels having the same spin as the ground state, and the
Case A assumption applies to lines of other multiplicities. In the
last of the line series of each ion, we present the adopted ionic
abundance and the error estimated from the line intensity-weighted
mean.  Effective recombination coefficients for the lines' parent
multiplet are the same as those used by Otsuka et al. (2010).  The RL
ionic abundances are insensitive to the electron density under
$\lesssim$10$^{8}$ cm$^{-3}$ (Zhang \& Liu 2003). For the ionic
abundance calculations, we therefore adopted the effective
recombination coefficients in case of {\Ne}=10$^{4}$ cm$^{-3}$ for
C$^{2+}$ and N$^{2+}$.  For He$^{+}$ and O$^{2+}$ calculations, we
adopted {\Ne}=10$^{6}$ cm$^{-3}$ and 10$^{4}$ cm$^{-3}$, respectively.
Since He~{\sc ii}, C~{\sc iii}, C~{\sc iv}, N~{\sc iii}, and O~{\sc
  iii} appeared to be of stellar origin, we did not estimate the
abundances of these ions.

The He$^{+}$ abundances are determined using electron density
insensitive six {\hei} lines to reduce intensity enhancement by
collisional excitation from the He$^{0}$ 2$s$ $^{3}S$ level.
For the C$^{2+}$ abundances, the V6 and V17.06 lines, which have
higher angular momentum as upper levels, seem to be unaffected by both
resonance fluorescence by starlight and recombination from excited
$^{2}S$ and $^{2}D$ terms.  Comparison of the C$^{2+}$ abundances
derived from V6 and V17.06 lines indicated that the observed C\,{\sc
  ii} lines would have less population enhancement mechanisms.  Our
estimated He$^{+}$ and C$^{2+}$ abundances are consistent with Henry
et al. (2010; 3.56(--2), 4.48(--4)).

We estimated the O$^{2+}$ abundances using the O~{\sc
ii} lines showing the least contamination from other ionic transitions. 
We excluded the O~{\sc ii} 4676.23
{\AA} line when determining O$^{2+}$ abundance, because this line is
much stronger than the other V1 lines.

The abundance discrepancy factor (ADF), which is the ratio of RL
  to CEL abundances, is 19.5$\pm$8.4 for O$^{2+}$ and 15.3$\pm$13.6
  for N$^{2+}$.  The large uncertainty of ADF(N$^{2+}$) is due to the
  large uncertainty in the CEL N$^{2+}$ abundance. ADF(O$^{2+}$) in
M1-11 is comparable to that of high density PN M2-24 (namely, 17;
Zhang \& Liu 2003). Zhang \& Liu (2003) also found that the density
structure for M1-11 and M2-24 is similar, with M2-24 showing a large
density contrast of {\Ne}=10$^{3-6}$ cm$^{-3}$.

\begin{deluxetable}{clccc}
\tablecolumns{5}
\tablecaption{Ionic abundances derived from RLs.\label{orl_abun}}
\tablewidth{240pt}
\tablehead{
\colhead{X$^{+}$}&
\colhead{Multi.}&
\colhead{$\lambda_{\rm lab}$}&
\colhead{$I$($\lambda$)}&
\colhead{X$^{m+}$/H$^{+}$}\\
\colhead{}&
\colhead{}&
\colhead{({\AA})}             &
\colhead{[$I$({\hb})=100]}&
\colhead{}
}
\startdata
He$^{+}$ & V11 & 5876.66 & 6.16(0) $\pm$ 8.32(--2) & 4.17(--2) $\pm$ 2.93(--3) \\
 & V14 & 4471.47 & 1.72(0) $\pm$ 3.02(--2) & 3.37(--2) $\pm$ 1.98(--3) \\
 & V45 & 7281.35 & 3.19(--1) $\pm$ 1.27(--2) & 3.60(--2) $\pm$ 2.74(--3) \\
 & V46 & 6678.16 & 1.60(0) $\pm$ 2.78(--2) & 3.83(--2) $\pm$ 1.97(--3) \\
 & V48 & 4921.93 & 4.64(--1) $\pm$ 7.25(--3) & 3.38(--2) $\pm$ 1.99(--3) \\
 & V51 & 4387.93 & 2.38(--1) $\pm$ 8.44(--3) & 3.78(--2) $\pm$ 2.99(--3) \\
 &  &  &     & {\bf 3.93(--2) $\pm$ 2.58(--3)} \\
C$^{2+}$ & V2 & 6578.01 & 4.99(--1) $\pm$ 1.25(--2) & 6.76(--4) $\pm$ 3.11(--4) \\
 & V3 & 7231.33 & 2.16(--1) $\pm$ 1.25(--2) & 5.13(--4) $\pm$ 2.26(--4) \\
 & V3 & 7236.42 & 4.36(--1) $\pm$ 1.08(--2) & 5.77(--4) $\pm$ 2.52(--4) \\
 & V3 & 7237.17 & 7.14(--2) $\pm$ 4.99(--3) & 8.44(--4) $\pm$ 3.74(--4) \\
 & V6 & 4267.15 & 6.14(--1) $\pm$ 2.68(--2) & 5.31(--4) $\pm$ 2.00(--4) \\
 & V17.06 & 5342.43 & 3.20(--2) $\pm$ 4.30(--3) & 5.37(--4) $\pm$ 2.42(--4) \\
 &  &  &    & {\bf 5.59(--4) $\pm$ 2.31(--4)} \\
N$^{2+}$ & V3 & 5666.63 & 1.01(--1) $\pm$ 4.80(--2) & 5.09(--4) $\pm$ 3.61(--4) \\
 & V3 & 5677.66 & 5.18(--2) $\pm$ 1.00(--2) & 5.51(--4) $\pm$ 3.09(--4) \\
 & V5 & 4630.54 & 5.49(--2) $\pm$ 7.05(--3) & 2.74(--4) $\pm$ 1.46(--4) \\
 & V19 & 5017.22 & 1.38(--2) $\pm$ 3.58(--3) & 5.12(--4) $\pm$ 3.07(--4) \\
   &  &  &  & {\bf 4.61(--4) $\pm$ 2.92(--4)} \\
O$^{2+}$ & V1 & 4641.81 & 2.61(--2) $\pm$ 1.01(--3) & 1.52(--4) $\pm$ 4.68(--5) \\
         & V1 & 4676.23 & 8.59(--2) $\pm$ 6.78(--3) & 9.22(--4) $\pm$ 1.19(--4) \\
         & V1 & 4649.14 & 2.20(--2) $\pm$ 6.00(--3) & 1.17(--4) $\pm$ 4.80(--5) \\
         & V1 & 4650.84 & 3.10(--2) $\pm$ 7.00(--3) & 2.00(--4) $\pm$ 7.59(--5) \\
 &  &    &  & {\bf 1.61(--4) $\pm$ 5.85(--5)}
\enddata
\end{deluxetable}

\subsection{Elemental abundances}
The elemental abundances are estimated using an ionization correction
factor, ICF(X), based on the IP. ICFs(X) for each element are listed
in Table \ref{ICF}.  The He abundance is the sum of the He$^{+}$ and
He$^{0+}$ abundances, and we allowed for the unseen He$^{0}$
abundance.  The C abundance is the sum of the C$^{+}$ and C$^{2+}$
abundances, and we added the unseen C$^{+}$ using ICF(C).  Henry
  et al. (2010) used ICF(C)=O/O$^{2+}$. Since the IPs of C$^{+,2+}$
  (11.3 and 24.4 eV) are close to those of N$^{+,2+}$ (14.5 and 29.6
  eV), we instead adopted ICF(C)=N/N$^{2+}$.  The N abundance is the
sum of N$^{+}$ and N$^{2+}$. For the RL N abundance, we corrected for
the unseen N$^{+}$ assuming (N/N$^{2+}$)$_{\rm CEL}$ =
(N/N$^{2+}$)$_{\rm RL}$. The O abundance is the sum of the O$^{+}$ and
O$^{2+}$ abundances.  For the RL O abundance, we assume
(O$^{2+}$/O)$_{\rm RLs}$ = (O$^{2+}$/O)$_{\rm CELs}$.  The Ne
abundance is equal to the Ne$^{+}$ abundance. The P abundance is the
sum of the P$^{+}$ and P$^{2+}$ abundances. We assumed that P/P$^{+}$
= S/S$^{+}$ and considered the P$^{2+}$ abundance. The S abundance is
the sum of the S$^{+}$, S$^{2+}$, and S$^{3+}$ abundances. We
corrected for the unseen S$^{3+}$ abundance using the CEL O and
O$^{+}$ abundances. We assume that the Cl abundance is the sum of
Cl$^{+}$ and Cl$^{2+}$.  The Ar abundance is the sum of the Ar$^{+}$
and Ar$^{2+}$ abundances, and we corrected for the unseen Ar$^{+}$.
The Fe abundance is the sum of the Fe$^{2+}$ and Fe$^{3+}$ abundances,
and we corrected for the unseen Fe$^{3+}$.  The Kr abundance is the
sum of the Kr$^{+}$, Kr$^{2+}$ and Kr$^{3+}$ abundances, and we
corrected for the unseen Kr$^{+}$ and Kr$^{3+}$.

The resulting elemental abundances are listed in Table \ref{abund}.
The types of emission lines used for the abundance estimations are
specified in the second column. The number densities of each element
relative to hydrogen are listed in the third column, and the
subsequent two columns are the number densities in the form of
$\log_{10}$(X/H), where H is 12, and the relative number densities to
the solar value. The last columns are the measurements by Sterling \&
Dinerstein (2008) for Kr and Henry et al. (2010) for the others. Our
estimated abundances are in agreement with the values given by these
authors except for C, N, and O. The C discrepancy between Henry et al.
(2010) and ours is due to the different ICF(C). When we assume
  ICF(C)=O/O$^{2+}$, we find that the C abundance is 10.45$\pm$0.22.
The N and O discrepancies are due to the N$^{+}$ and O$^{+}$
discrepancies caused by the differently adopted {\te}'s.

\begin{deluxetable}{@{}llll@{}}
\tablecolumns{4}
\tablecaption{Adopted Ionization Correction Factors (ICFs).}
\tablewidth{240pt}
\tablehead{
\colhead{X}&
\colhead{Type}&
\colhead{ICF(X)}&
\colhead{X/H}
}
\startdata
He &RL &S/S$^{2+}$&ICF(He)He$^{+}$\\
%
C  &RL &(N/N$^{2+}$)$_{\rm CEL}$&ICF(C)C$^{2+}$\\
%
N &CEL &1 &ICF(N)(N$^{+}$+N$^{2+}$)\\
  &RL &(N/N$^{2+}$)$_{\rm CEL}$&ICF(N)N$^{2+}$\\
%
O  &CEL & 1 &ICF(O)(O$^{+}$+O$^{2+}$)\\
   &RL &(O/O$^{2+}$)$_{\rm CEL}$ &ICF(O)O$^{2+}$\\
Ne &CEL & 1 &ICF(Ne)Ne$^{+}$\\
%
P &CEL &(S/S$^{+}$)&ICF(P)P$^{+}$\\

S &CEL &$\left[1 - (1-({\rm O^{+}/O}))^{3}\right]^{-1/3}$ &ICF(S)(S$^{+}$+S$^{2+}$)\\

Cl &CEL &1 &ICF(Cl)$\rm \left(Cl^{+}+Cl^{2+}\right)$\\

Ar &CEL &(N/N$^{2+}$)$_{\rm CEL}$&ICF(Ar)Ar$^{2+}$\\

Fe &CEL &(O/O$^{+}$)&ICF(Fe)Fe$^{2+}$\\

Kr &CEL &Cl/Cl$^{2+}$ &ICF(Kr)Kr$^{2+}$
\enddata
\label{ICF}
\end{deluxetable}

\begin{deluxetable}{@{}cccrc|r@{}}
\tablecolumns{6}
\tablecaption{Elemental abundances of M1-11.}
\tablewidth{240pt}
\tablehead{
\colhead{X} &
\colhead{Type} &
\colhead{X/H} &
\colhead{$\log$(X/H)+12}   &
\colhead{[X/H]$^{a}$}   &
\colhead{Others}
}
\startdata
He & RL & 5.65(--2)$\pm$7.36(--3) & 10.75$\pm$0.06 & --0.15$\pm$0.06 & 10.55 \\
C & RL  & 2.19(--3)$\pm$1.60(--3) & 9.34$\pm$0.32 & +0.95$\pm$0.41 & 9.78 \\
N & CEL  & 1.22(--4)$\pm$1.85(--5) & 8.08$\pm$0.07 & +0.25$\pm$0.13 & 7.67 \\
 & RL   & 1.81(--3)$\pm$1.58(--3) & 9.26$\pm$0.38 & +1.43$\pm$0.60 & \nodata \\
O & CEL  & 4.13(--4)$\pm$2.40(--5) & 8.62$\pm$0.03 & --0.07$\pm$0.06 & 7.88 \\
 & RL   & 8.05(--3)$\pm$3.45(--3) & 9.91$\pm$0.19 & +1.22$\pm$0.21 & \nodata \\
Ne & CEL & 4.76(--5)$\pm$2.98(--6) & 7.68$\pm$0.03 & --0.19$\pm$0.10 & \nodata \\
P & CEL  & 7.51(--8)$\pm$8.59(--9) & 4.88$\pm$0.06 & --0.58$\pm$0.06 & \nodata \\
S & CEL  & 1.43(--6)$\pm$9.34(--8) & 6.15$\pm$0.03 & --1.04$\pm$0.05 & 5.96 \\
Cl & CEL & 2.83(--8)$\pm$5.56(--9) & 4.45$\pm$0.09 & --0.81$\pm$0.11 & 4.11 \\
Ar & CEL & 9.85(--7)$\pm$6.15(--7) & 5.99$\pm$0.27 & --0.81$\pm$0.11 & 5.37 \\
Fe & CEL & 1.13(--7)$\pm$2.97(--8) & 5.05$\pm$0.11 & --2.42$\pm$0.12 & \nodata \\
Kr & CEL & 6.48(--9)$\pm$2.21(--9) & 3.81$\pm$0.15 & +0.53$\pm$0.17 & 4.25
\enddata
\tablenotetext{a}{The solar abundances are from Lodders (2003).}
\label{abund}
\end{deluxetable}

\section{Discussion}

\subsection{C and O abundances and the C/O abundance ratio}
We attempted to estimate the carbon abundance from CELs by correlating
ADF(C$^{2+}$) with ADF(O$^{2+}$). We found a tight correlation,
ADF(C$^{2+}$)$=0.997\times$ADF(O$^{2+}$), among 56 Galactic PNe from
Wang \& Liu (2007), Wesson et al. (2005), Liu et al. (2004), and
Tsamis et al. (2004). When we consider only the 18 PNe with
ADF(O$^{2+}$)$>$5, there seems to be no correlation between
ADF(C$^{2+}$) and ADF(O$^{2+}$). For these 18 PNe ADF(C$^{2+}$) is
7.65$\pm$1.97. When we adopt this value for ADF(C$^{2+}$), the
C$^{2+}$ abundance derived from CELs is 7.31(--5)$\pm$3.56(--5) and
the $I$(C~{\sc iii}]$\lambda$1906/09) is 33.8$\pm$16.5.  The CEL C
abundance and the CEL C/O ratio are 8.46$\pm$0.34 dex and
--0.14$\pm$0.34 dex, respectively. The CEL C/O ratio derived from CELS
agrees well with the value derived from RLs, which is --0.57$\pm$0.37
dex, although there is a large uncertainty.

We should keep in mind that the $>$9 dex C abundances derived from
both the CELs and the RLs are confirmed in many PNe and can also be
theoretically explained by AGB nucleosynthesis if we assume a small
minimum H-envelope mass for the third dredge-up ($\sim$0.5
$M_{\odot}$; Straniero et al. 1997 for solar metallicity). However,
for stars with a main sequence mass of $\sim$1-3 $M_{\odot}$ and an
initial metallicity of $Z$=0.004, O abundances of $>$9 dex are not
expected to occur.  For example, we present the predicted abundances
for initially 1.5 $M_{\odot}$ stars with in Table \ref{modelabun}. In
M1-11, we adopt the observed N and O abundances derived from CELs,
while for the C abundances we used CEL {\it predictions} and
observed values for the RLs.  Karakas (2010) adopted scaled-solar
abundances as the initial composition. The accuracy of the predicted
abundances is within 0.3 dex. The model predicted C and O abundances
are close to the {\it predicted} CEL C and the observed CEL O
abundances. The model could also explain the observed abundance of Ne
and P, which are He-rich intershell products that are brought up to
the stellar surface by the third dredge-up in late AGB phase.

Comparison of the model of Karakas (2010) and our observation
suggests that the CEL O abundance would be more reliable in M1-11 
relative to the RL O. This would be also applied to the N abundance. 
Therefore, we consider  the CEL CNO abundances to represent the
gas-phase abundances of these elements in M1-11. If the CEL C/O 
ratio is correct, M1-11 might be a C-rich or O-rich PN. In the next 
section, we verify the C and O abundances through gas+dust SED modeling.

\begin{deluxetable}{@{}ccccccl@{}}
\tablecolumns{7}
\tablecaption{Comparison of the observed abundances with the prediction
 by an initially 1.5 $M_{\odot}$ star with $Z$=0.004 (see text in detail). \label{modelabun}}
\tablewidth{240pt}
\tablehead{
\colhead{He}&
\colhead{C}&
\colhead{N}&
\colhead{O}&
\colhead{Ne}&
\colhead{P}&
\colhead{Sources}
}
\startdata
10.75 &9.34/8.46&8.08&8.62&7.68&4.88&This work\\
10.97 &8.46     &7.65&8.23&7.42&4.86&Karakas (2010)
\enddata
\end{deluxetable}

\subsection{SED modeling}
We investigate the fractional ionization, the evolutionary stage of the
central star,  and the physical
conditions of the ionized gas and dust grains, by constructing an SED model
based on the photo-ionization (P-I) code {\sc Cloudy} c08.00 and
modified blackbody fitting. Through SED modeling, we derive the
ionized gas mass $m_{g}$, dust mass $m_{d}$, and dust temperature $T_{d}$
in the nebula, as well as the total luminosity $L_{\ast}$ and the effective stellar
temperature $T_{\rm eff}$. We verify CEL C and O abundances through
our model.

\subsubsection{Model approach}
Using {\sc Cloudy} and modified blackbody fitting, we attempted to fit the
observed SED from $\sim$0.1 to 90 $\mu$m, assuming that the dust in
M1-11 is composed of PAHs, amorphous carbon (AC), SiC, and MgS grains. 
We excluded the broad 16-22 $\mu$m feature in the fitting procedure.

Since there are no optical constants of MgS available for the UV, we
could not include MgS in the {\sc Cloudy} modeling.  Instead, we
fitted the MgS feature using single temperature thermal emission
applied to the infrared opacity.  For MgS, we assumed CDE, while for
all other dust species we used spherical grains.  Unfortunately, {\sc
  Cloudy} does not allow CDE grains, providing a second reason to
model the MgS contribution separately.  Indeed, at present, there are
no publicly available P-I codes which can handle CDE grain shapes.

We assume that the observed SEDs can be expressed as the sum of the
calculated SED from the {\sc Cloudy} model and the modified blackbody
fit to the MgS emission. First, we fitted the SED in the range from
$\sim$0.1 to 22.5 $\mu$m and 65-90 $\mu$m using {\sc Cloudy}. Then, we
fitted a modified blackbody to the difference between the observations
and the fitted SED from {\sc Cloudy}, to account for the contribution
from MgS.  In both the {\sc Cloudy} model and the MgS modified
blackbody fitting, we adopted a distance $D$ to M1-11 of 2.1 kpc.

In the {\sc Cloudy} calculations, we used Tlusty's non-LTE theoretical
atmosphere
model\footnote[10]{http://nova.astro.umd.edu/Tlusty2002/tlusty-frames-cloudy.html}
with [Z/H]=--0.81, which is consistent with the observed [Ar/H].
McCarthy et al. (1997) found that the effective temperature $T_{\rm
  eff}$=29\,000 K, the surface gravity $\log$~$g$=3.0, and the core mass
$M_{\ast}$=0.74 $M_{\odot}$ for the central star, using their non-LTE
model based on Keck/HIRES spectra. However, they adopted a distance
$D$ of 4.6 kpc. Guided by their results, we used a series of
theoretical atmosphere models with $T_{\rm eff}$ in the range from
27\,500 to 35\,000 K and $\log$~${g}$ values of 3.0, 3.25, 3.50, 3.75,
and 4.0 to describe the SED of the central star.  We found that
$T_{\rm eff}$=31\,950 K, $\log$~${g}$=3.5, and a total luminosity
$L_{\ast}$=4510 $L_{\odot}$ can accommodate the observations well. 

For the gas-phase elemental abundances X/H, we adopted the
  observed values listed in Table \ref{abund} as a first guess and
  varied these to match observations, except for C. For C, we varied
  the C abundance to match the predicted value: $I$(C~{\sc
    iii}]$\lambda$1906/09)=33.8$\pm$16.5.  We did not fit the He~{\sc
  ii}, C~{\sc iii} or N~{\sc iii} lines, because these lines would be
of stellar wind origin. For the N abundances, we adopted the CEL
value. Treating the collisional impacts of [P\,{\sc ii}] as a function
of electron temperature and solving the 5-level energy model for this
ion is not included in {\sc Cloudy} c08.00, leading to an overestimate
of the P abundance ($\log_{10}$P/H + 12 $>$6). We adapted the code to
perform this analysis correctly, and obtain a better estimate of the P
abundance. We also revised the C~{\sc iii}], [N~{\sc ii}],
  [O~{\sc ii, iii}], [Ne~{\sc ii}], [S~{\sc iii}], and [Ar~{\sc iii}]
  line-calculation programs in {\sc Cloudy}.  The abundance of
elements that were not observed was fixed at [X/H]=--0.81. Reliable
dielectric recombination (DR) rate measurements do not exist for low
stages of ionization of S at photo-ionization temperatures ({\sc
  Cloudy} c08 manual), thus we adopted the scaled DR of oxygen for
sulfur line calculations, to match the observed {\sii}.

We adopted the hydrogen density profile as presented in
Fig.~\ref{radial}(b). Based on the \emph{HST} image shown in
Fig.~\ref{image}, we fixed the outer radius $R_{\rm out}$ at 2.6$''$
(0.026 pc) and varied the inner radius $R_{\rm in}$ and filling factor
$\epsilon$ to match the observed SED. The best fit $R_{\rm in}$ and
$\epsilon$ are $\sim$0.8$''$ (0.008 pc) and 0.25-0.3, respectively,
assuming that the gas and dust co-exist in the same sized nebula.

We assumed that amorphous carbon and SiC are
present in the form of spherical dust grains with a radius $a$=0.5
$\mu$m.  To calculate the opacities for these species, we used the
optical constants listed in Table~\ref{optdata}.  Zhang \& Kwok (1990)
observed a near IR-excess, also shown in the SED of Fig.~\ref{sed},
arising from both the PAH and the small dust grain emission. To
explain the excess, we also included small carbonaceous grains with a
grain size of $a$=0.005 $\mu$m, as well as PAHs.  We adopted the
optical data of Desert et al. (1990),  Schutte et al. (1993), and 
Bregman et al. (1989). To explain the MgS
feature around 30 $\mu$m, we performed a modified blackbody fitting
with a single dust temperature $T_{d}$ and CDE shaped grains with a
characteristic size of $a$=0.5 $\mu$m as we mentioned in Section
2.6.2.

To verify the degree of modeling accuracy, we evaluated the gas
emission line strengths and the broad band fluxes in the features
of interest (Table \ref{cloudy}), including three \emph{AKARI/FIS}
and two Wide-field Infrared Survey
Explorer (\emph{WISE}) photometry bands; and we fitted the overall SED shape
with the adopted elemental abundances. We will discuss the resulting
elemental abundance and the ionization correction factors.

\subsubsection{Results of SED modeling}

The derived physical quantities are listed in Tables~\ref{cloudypara2}
and \ref{cloudy}, which can be compared with the flux levels in
the 13 spectral features of interest. In Table~\ref{cloudy}, we
compare the predicted emission line and band fluxes with the observed
values, where $I$({\hb}) is 100 and the intrinsic $\log$~$I$({\hb}) is
--10.22 erg s$^{-1}$ cm$^{-2}$. Columns 4 and 5 list the observed
values and the values predicted by {\sc Cloudy} modeling. The type of
emission is indicated in column 3.

The predicted CEL C and O abundances are close to the model results
by Karakas (2010). The predicted CEL C/O ratio (+0.19 dex)
suggests that M1-11 could be a normal low-mass C-rich PN slowly 
evolving toward higher effective temperatures. The elemental
abundances predicted by {\sc Cloudy} are comparable with the observation
and are in excellent agreement with the model prediction by Karakas (2010).

\begin{figure}[t]
\epsscale{1.0}
\plotone{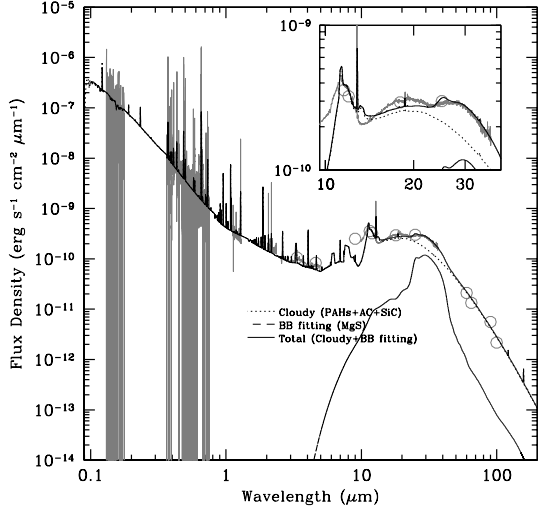}
\caption{The fitted SED from the {\sc Cloudy} modeling (dots) and
 the modified blackbody fitting (long dash) and the resultant SED (thick black
  line). The gray circles and lines are data from \emph{IUE}, Subaru/HDS,
  OAO/ISLE spectra, 2MASS (Ramos-Larios \& Phillips 2005), \emph{AKARI}/IRC
  spectra and 4 bands (9/18/65/90 $\mu$m), and \emph{Spitzer}/IRS spectrum.
  In the {\sc Cloudy} model, we considered PAHs, amorphous carbon
  (AC), and SiC. The modified blackbody fitting is
  performed for MgS only. The close-up feature of the observed and
  fitted SEDs around 10-40 $\mu$m are presented in the inner box.  See
  main text for a detailed description. \label{sed}}
\end{figure}

In Fig.~\ref{sed}, we present the calculated SED from {\sc Cloudy}
(dots), the MgS modified blackbody fitting (broken line), and the sum of those
two components (thick line). The gray lines and circles represent the
observations including the \emph{IUE} spectra, 2MASS $JHKs$
(Ramos-Larios \& Phillips 2005), the \emph{AKARI}
9/18/65/90 $\mu$m, and the \emph{WISE} 3 and 4 $\mu$m photometry. In the inset, we focus on the
$\sim$10-40 $\mu$m part of the SED. The predicted SED matches the data in UV to
mid-IR range well, except for the PAH emission, in particular the
13.6$\mu$m feature due to the C-H out of plane bending mode (quartet).

The temperatures of the 0.5-$\mu$m sized amorphous carbon, SiC, and MgS
are  91, 102 (by {\sc Cloudy}), and 160-200 K (by blackbody fitting), 
respectively. The temperature of the 0.005-$\mu$m sized 
amorphous carbon grains and PAHs are 177 K and 267 K, respectively. 
The gas C mass within 2.6$''$ predicted by {\sc Cloudy} is 
4.68(--5) $M_{\odot}$. Most of the C in the nebula is existing as the grains. 
The mass of pure C$_{60}$ in the nebula 
is only $\sim$0.008 $\%$ of the total C dust.

\begin{figure}
\epsscale{1.1}
\plotone{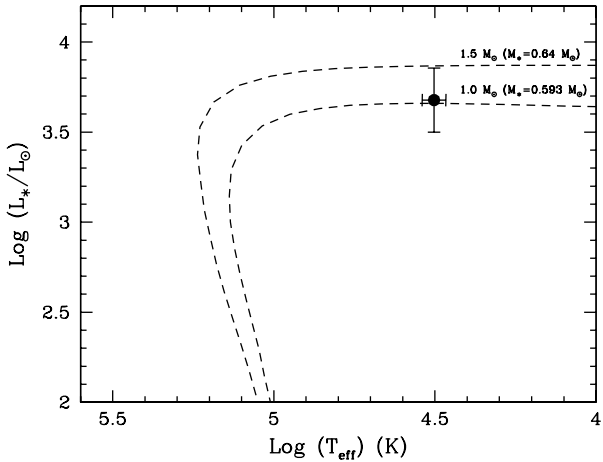}
\caption{The location of M1-11 on the 1.0 and 1.5 $M_{\odot}$
hydrogen burning evolutionary tracks after AGB phase. The location of
 M1-11 is indicated by
the filled circle.\label{hr}}
\end{figure}

Assuming a distance of 2.1 kpc,
the initial mass of the progenitor is estimated to be 1-1.5
$M_{\odot}$ based on the location of M1-11 with respect to the post-AGB
H-burning evolutionary tracks with $Z$ = 0.004 (Vassiliadis \& Wood
1994), as shown in Fig.~\ref{hr}. The age is $>$1000 yrs after leaving
the AGB phase.

\begin{deluxetable}{@{}lc@{}}
\tablecolumns{2}
\tablecaption{The derived properties of the central star, ionized nebula, and dust by
 the SED model. \label{cloudypara2}}
\tablewidth{240pt}
\tablehead{
\colhead{Parameters}      &
\colhead{Central Star}
}
\startdata
$L_{\ast}$        &4710 $L_{\odot}$\\
$T_{\rm eff}$ &31\,830 K        \\
$\log\,g$ &3.5 cm s$^{-2}$\\
$[Z]$ & --0.81\\
$M_{\ast}$   &$\sim$0.6 $M_{\odot}$\\
Progenitor   &1-1.5 $M_{\odot}$\\
Age          &$>$1000 yr\\
Distance  &2.1 kpc\\
\hline
      &Nebula\\
\hline
Abundances      &He:11.11,C:8.49,N:7.89,O:8.30,\\
($\log$X/H+12) &Ne:7.67,P:5.21,S:6.15,Cl:4.20,\\
                &Ar:5.95,Fe:5.01,Others:[X/H]=--0.81\\
Geometry        &Spherical\\
Shell size      &$R_{\rm in}$=0.8$''$(0.008 pc)/$R_{\rm out}$=2.6$''$(0.027 pc)\\
input density profile &See Fig.\ref{radial}(b)\\
$\epsilon$&0.25\\
$\log F$({\hb}) &--10.22 erg s$^{-1}$ cm$^{-2}$ (de-redden)\\
$m_{\rm g}$({\sc Cloudy})&0.023 $M_{\odot}$\\
\hline
      &Dust in Nebula\\
\hline
Composition &PAHs,AC,SiC,MgS\\
Grain size               &0.5 \& 0.005 $\mu$m (See text in detail)\\
Grain shape &CDE for MgS, spherical for the others\\
$T_{d}$({\sc Cloudy})$^{a}$    &90--267 K (See text)\\
$T_{d}$({\sc BB fitting})$^{b}$  &160--200 K\\
$m_{d}$({\sc Cloudy})$^{a}$    &3.44(--4) $M_{\odot}$\\
$m_{d}$({\sc BB fitting})$^{b}$ &4.81(--6)--9.27(--6) $M_{\odot}$ \\
$m_{d}$({\sc Tot.})$^{c}$      &3.49(--4)--3.53(--4) $M_{\odot}$ \\
$m_{d}$/$m_{\rm g}$({\sc Cloudy})&1.52(--2)
\enddata
\tablecomments{The accuracy of $\log$(X/H) + 12 is 0.3 dex. AC stands for amorphous carbon.}
\tablenotetext{a}{The temperature and total mass of PAHs, AC, and SiC grains derived by {\sc Cloudy}.}
\tablenotetext{b}{The temperature and mass of MgS grains derived
by the modified blackbody fitting.}
\tablenotetext{c}{The total mass of PAHs, AC, SiC, and MgS grains.}
\end{deluxetable}

\begin{deluxetable}{@{}cccrr@{}}
\tablecolumns{5}
\tablecaption{The predicted relative fluxes by the {\sc Cloudy} models.\label{cloudy}}
\tablewidth{240pt}
\tablehead{
\colhead{Ions}	&	
\colhead{$\lambda_{\rm lab}$}		&	
\colhead{Type}	&	
\colhead{$I$($\lambda$)$_{\rm obs}$}	&	
\colhead{$I$($\lambda$)$_{\rm cloudy}$}	
}
\startdata
He$^{+}$&	4387.93	{\AA}	&	RL	&	2.38(--1)	&	2.69(--1)	\\
	&	4471.47	{\AA}	&	RL	&	1.72(0)	&	2.22(0)	\\
	&	4921.93	{\AA}	&	RL	&	4.64(--1)	&	5.84(--1)	\\
	&	5876.66	{\AA}	&	RL	&	6.16(0)	&	6.76(0)	\\
	&	6678.16	{\AA}	&	RL	&	1.60(0)	&	1.74(0)	\\
	&	7281.35	{\AA}	&	RL	&	3.19(--1)	&	4.70(--1)	\\
C$^{2+}$	&	1906/09	{\AA}	&	CEL	&$<$5.03(+1)	&	4.52(+1)	\\
N$^{+}$	&	5754.64	{\AA}	&	CEL	&	5.87(0)	&	6.92(0)	\\
	&	6548.04	{\AA}	&	CEL	&	5.89(+1)	&	6.77(+1)	\\
	&	6583.46	{\AA}	&	CEL	&	1.90(+2)	&	2.00(+2)	\\
N$^{2+}$&	1750    {\AA}	&	CEL	&	1.41(0)	&	1.04(0)	\\
O$^{+}$	&	3726.03	{\AA}	&	CEL	&	5.54(+1)	&	6.97(+1)	\\
	&	3728.81	{\AA}	&	CEL	&	1.97(+1)	&	2.94(+1)	\\
	&	7320/30	{\AA}	&	CEL	&	2.58(+1)	&	4.39(+1)	\\
O$^{2+}$&       4361.21 {\AA}	&	CEL	&	1.37(--1)	&	1.80(--1)	\\
	&	4958.91	{\AA}	&	CEL	&	6.44(0)	&	6.07(0)	\\
	&	5006.84	{\AA}	&	CEL	&	1.85(+1)	&	1.83(+1)	\\
Ne$^{+}$	&12.81 $\mu$m	&	CEL	&	3.09(+1)	&	3.18(+1)	\\
P$^{+}$	&	1.15 $\mu$m	&	CEL	&	1.22(--1)	&	1.31(--1)	\\
	&	1.19 $\mu$m	&	CEL	&	3.41(--1)	&	3.93(--1)	\\
S$^{+}$	&	4068.60	{\AA}	&	CEL	&	1.79(0)	&	1.27(0)	\\
	&	4076.35	{\AA}	&	CEL	&	6.34(--1)	&	 4.07(--1)	\\
 	&	6716.44	{\AA}	&	CEL	&	4.42(--1)	&	4.56(--1)	\\
	&	6730.81	{\AA}	&	CEL	&	1.08(0)	&	9.08(--1)	\\
S$^{2+}$	&6312.10 {\AA}	&	CEL	&	3.23(--1)	&	5.50(--1)	\\
	&	9068.60	{\AA}	&	CEL	&	3.22(0)	&	3.89(0)	\\
	&	18.71 $\mu$m	&	CEL	&	1.40(0)	&	3.65(0)	\\
Cl$^{+}$	&8578.69 {\AA}	&	CEL	&	1.19(--1)	&	5.94(--2)	\\
	&	9123.60	{\AA}	&	CEL	&	7.76(--2)	&	1.57(--2)	\\
Cl$^{2+}$	&5517.66 {\AA}	&	CEL	&	2.88(--2)	&	3.24(--2)	\\
	&	5537.60	{\AA}	&	CEL	&	7.22(--2)	&	7.62(--2)	\\
Ar$^{2+}$	&5191.82 {\AA}	&	CEL	&	1.86(--2)	&	2.62(--2)	\\
	&	7135.80 {\AA}	&	CEL	&	2.46(0)	        &	3.04(0)	\\
        &	8.99 $\mu$m	&	CEL	&	3.06(0)	        &	1.91(0)	\\
Fe$^{2+}$	&4701.53{\AA}	&	CEL	&	6.89(--2)	&	6.62(--2)	\\
	&	4754.69	{\AA}	&	CEL	&	3.86(--2)	&2.70(--2)	\\
	&	4769.43	{\AA}	&	CEL	&	5.32(--2)	&	2.22(--2)	\\
	&	4881.00	{\AA}	&	CEL	&	4.93(--2)	&7.07(--2)	\\
        &       5270.40	{\AA}	&	CEL	&	7.06(--2)       &8.85(--2)\\
\hline
\smallskip
Bands    & $\lambda_{center}$          &$\Delta\lambda$    & $I$($\lambda$)$_{\rm
 obs}$&$I$($\lambda$)$_{\rm cloudy}$\\
\hline
$B$      & 0.40 $\mu$m &0.300 $\mu$m &4.05(+2)  & 4.36(+2) \\
$J$      & 1.24 $\mu$m &0.162 $\mu$m &1.14(+2)  & 1.24(+2) \\
$H$      & 1.66 $\mu$m &0.251 $\mu$m &9.21(+1)  & 1.12(+2)   \\
$Ks$      & 2.16 $\mu$m&0.262 $\mu$m &8.36(+1)  & 8.01(+1)    \\
$WISE1$  & 3.353 $\mu$m & 0.663 $\mu$m&1.15(+2)  & 1.32(+1)   \\
$WISE2$  & 4.603 $\mu$m & 1.042 $\mu$m&1.43(+2)  & 1.27(+2)   \\
$AKARI1$ & 9.22 $\mu$m  &4.104 $\mu$m &1.62(+3)  & 1.01(+3) \\
$IRS1$   &10.30 $\mu$m  &1.000 $\mu$m &  2.21(+2)& 2.18(+2)\\
$IRS2$   & 11.30 $\mu$m &1.000 $\mu$m &6.54(+2)  & 6.81(+2) \\
$IRS3$   & 12.50 $\mu$m &1.000 $\mu$m &5.14(+2)  & 5.14(+2) \\
$IRS4$   & 13.50 $\mu$m &0.600 $\mu$m   &2.08(+2)  &2.56(+2)\\
$AKARI2$ &66.70 $\mu$m  & 20.2 $\mu$m &4.49(+2)  &5.04(+2)\\
$AKARI3$ &89.20 $\mu$m  & 39.9 $\mu$m &3.70(+2)  &3.47(+2)
\enddata
\end{deluxetable}

\subsection{Fullerene formation in PNe}

\begin{deluxetable*}
{@{}l@{\hspace{-2pt}}c@{\hspace{-2pt}}c@{\hspace{-2pt}}c@{\hspace{-2pt}}c@{\hspace{-2pt}}c@{\hspace{-2pt}}c@{\hspace{-2pt}}r@{\hspace{2pt}}c@{\hspace{-2pt}}c@{\hspace{2pt}}l@{}}
\tablecolumns{11}
\tablecaption{Physical parameters of C$_{60}$ detected Galactic PNe.\label{c60}}
\tablewidth{0pt}
\tablehead{
\colhead{Nebula} &
\colhead{$T_{\rm eff}$}&
\colhead{$L_{\ast}$}   &
\colhead{$\log~{g}$}   &
\colhead{$M_{prop.}$$^{a}$}  &
\colhead{Age$^{a}$}&
\colhead{Type of}      &
\colhead{Gas-phase abundances$^{b}$}&
\colhead{Dust}              &
\colhead{$T_{C60}$}              &
\colhead{Ref.}\\
\colhead{}       &
\colhead{(K)}      &
\colhead{($L_{\odot}$)}&
\colhead{(cm s$^{-2}$)}&
\colhead{($M_{\odot}$)}&
\colhead{(yrs)}&
\colhead{Central star} &
\colhead{(He/C/N/O/Ne/ )}&
\colhead{}  \\
\colhead{}       &
\colhead{ }      &
\colhead{ }&
\colhead{ }&
\colhead{}&
\colhead{ }&
\colhead{ } &
\colhead{(P/S/Cl/Ar/Kr)}&
\colhead{}&
\colhead{(K)}
}
\startdata
M1-11  & 31\,950   &4510&3.5&1-1.5&$>$1000    &$[$WC10-11$]$ &10.75/9.34/8.08/8.62/7.68/ &~~PAHs,SiC?,AC,MgS?&399&~~(1),(2),(3)\\
     &  &   &   &      &     &                      &  4.88/6.15/4.45/5.66/3.81  &   &  \\
M1-12  & 33\,000   &8700&3.4&$\sim$1.5&$\sim$1000&$[$WC10-11$]$&10.54/9.40/7.55/8.11/7.78/&~~PAHs,SiC?,AC,MgS?&345&~~(1),(2),(3),(4),(5)\\
     &   &   &   &      &     &                      &  \nodata/6.25/5.01/5.40/3.83  &   &  \\
Tc1    &34\,700    &6900&3.4&$\sim$2&$<$1000     &Of(H)
 &$>$10.78/8.56/7.56/8.42/7.80/&~~(weak)PAHs?,AC,MgS?&339&~~(1),(4),(6),(7),(8),(9)\\
      &   &  &   &      &    &                       & 5.30/6.45/4.97/6.71/\nodata  &   & \\
M1-20  &53\,000    &8300&5.0&1-1.5&$<$3900       &wel                       &10.98/8.74/7.94/8.58/8.21/&~~PAHs?,SiC?,AC,MgS?&425&~~(1),(4),(7),(10),(11),(12)  \\
      &   &  &   &      &    &                       & \nodata/6.68/4.93/5.85/\nodata  &   &
\enddata
\tablenotetext{a}{The values are estimated from the location on the
 H-burning tracks with $Z$=0.02 for Tc1 and $Z$=0.004
 for the others by Vassiliadis \& Wood (1994).}
\tablenotetext{b}{The number density relative to the hydrogen is defined
 as $\log$~H=12. The C abundance is derived from RLs except for
 Tc1. The C in Tc1 is derived from CEL. The N, O, and Ne abundances are
 derived from CELs. The predicted abundances in M1-11 by {\sc Cloudy}
 are listed in Table \ref{cloudypara2}.}
\tablerefs{(1) This work; (2) Weidmann \& Gamen (2011a); (3) Sterling \&
 Dinerstein 2008; (4) Zhang \& Kwok (1993); (5) Henry et al. (2010);
(6) Cami et al. (2010); (7) Garc{\'{\i}}a-Hern{\'a}ndez et al. (2010); (8)
 Pottasch et al. (2011); (9) Weidmann \& (2011b); (10) Wang
 \& Liu (2007); (11) G{\'o}rny et al. (2009); (12) Kaler \& Jacoby (1991)}
\end{deluxetable*}

Since the discovery of fullerenes C$_{60}$ and C$_{70}$ in a C-rich PN Tc1 
(Cami et al. 2010), the number of detections of these lines 
are increasing (Garc\'{i}a-Hern\'{a}ndez et
al. 2010, 2011a; Zhang \& Kwok 2011); however, the formation process of fullerenes in 
evolved stars and in the ISM is under debate. At present three explanations 
for fullerene formation are proposed.

\subsubsection{Destruction of HAC}
Garc\'{i}a-Hern\'{a}ndez et al. (2010) suggested that PAHs and
fullerenes may be formed by the photochemical processing of HAC in
H-rich circumstellar envelopes. In laboratory experiments, Scott et
al. (1997) showed that PAHs and C$_{50}$ ,C$_{60}$, and C$_{70}$ may be
produced by the decomposition of HACs. This fullerene formation scenario
seems to be supported by the \emph{Spitzer} observations of H-poor R
Coronae Borealis (RCB) stars (Garc\'{i}a-Hern\'{a}ndez et al. 2011b) In particular, the mid-IR spectrum of the R Coronae Borealis
star V854 Cen - an H-rich RCB star - evolved from HACs
(\emph{ISO}/SWS spectrum in 1996) to PAHs and C$_{60}$ (\emph{Spitzer}/IRS in
2007). The \emph{ISO}/SWS spectrum of this object is similar to a laboratory
spectrum of HAC at 773 K (Lambert et al. 2001). Later,
Garc\'{i}a-Hern\'{a}ndez et al. (2011a) show that seven Magellanic Cloud
PNe have broad features around 6-9 and 10-14 $\mu$m. The central
stars of these PNe are H-rich. The 6-9 $\mu$m broad feature is different
from PAH features in the same wavelength range and and is rather similar
to that of HACs. These MC PNe have cool central stars (31\,300-43\,300 K,
meaning young PNe), and their UV spectra, except for SMP SMC13 and LMC2,
show P-Cygni profiles, suggesting the presence of a stellar wind. This
suggests that the shocks may be triggering the HAC decomposition and
forming both fullerenes and PAHs. More recently, in the strong 
C$_{60}$ PNe Tc1, SMC16, and LMC56, which are showing HAC features,  
Bernard-Salas et al. (2012) argued that the UV radiation from the
central star is photochemically processing the HACs to produce
fullerenes, by comparing between the observed intensity ratios of
C$_{60}$ 7.0, 8.5, 17.3 to 18.9 $\mu$m and the theoretical predictions 
(see their Figs. 5 and 6). In Tc1, the intensity peak of C$_{60}$ 8.5
$\mu$m is 6400-9700 AU from the central star, while that of dust components is 
nearby the central star. It would be interesting that the intensity of 
the 11.2 $\mu$m feature peaks $\sim$3000 AU away from the central star 
on the other side of of C$_{60}$ 8.5 $\mu$m (see their Fig.3), 
because HACs seem to have a spectral peak around 11.2 $\mu$m (Jones 2012).

\subsubsection{Destruction of PAH clusters} 
Cami et al. (2010) argued that fullerenes are not formed from HACs, because if fullerenes
formed by HAC decomposition process, 
fullerenes should be observed more frequently in objects with
PAHs. 
Cami et al. (2011) suggested that fullerenes are products of 
the destruction of large PAHs ($>$60 C
atoms) by the stellar wind shock ($\sim$100 km s$^{-1}$). PAHs first
lose their peripheral H-atoms in the shock and in the post shock gas, 
and then the resulting carbon clusters can assemble into fullerenes.

\subsubsection{Photo-chemical process of PAH clusters}
In the reflection nebula NGC 7023, Sellgren et al. (2010) show that 
the C$_{60}$ 18.9 $\mu$m emission peaks on the central star, while 
the PAH 16.4 $\mu$m emission is brightest between the regions of strong 18.9 $\mu$m 
and the H$_{2}$ 0-0S(1) 17 $\mu$m emission.
Bern\'{e} \& Tielens (2012) found that the abundance of C$_{60}$ and PAHs 
in this nebula increases and decreases, respectively, when approaching 
the ionizing star HD200775 ($T_{\rm eff}$$\sim$19\,000 
K, e.g.  Alecian et al. 2008). They proposed that fullerenes are
products of the photo-chemical processing of large PAHs.
Such large PAH clusters could be from HACs as described above.

\subsubsection{Comparison between C$_{60}$ PNe}
Which explanations are more suitable to describe fullerene formation in
Galactic PNe? To answer this question, we checked the properties
of the central stars and the dust components, as summarized in Table
\ref{c60}. Garc\'{i}a-Hern\'{a}ndez et
al. (2010) detected fullerenes in K3-54; however, the properties of 
the central star and gas-phase elemental abundances are unknown. Therefore, we do not list this PN.
The progenitor masses $M_{prop.}$ and the evolutionary  age
after AGB phase in the fifth and sixth columns are estimated
from the location on the H-burning tracks of Vassiliadis \& Wood (1994)
with $Z$=0.02 for Tc1 and $Z$=0.004 for the others. 
Prior to our study, The C and Ne abundances in 
M1-12 were not well known. However, using the measured $T_{\rm
e}$([S~{\sc iii}]) and $n_{\rm e}$([S~{\sc ii}]), the intensities 
of the C~{\sc ii} 6462 {\AA}, and the de-reddened H${\beta}$ flux given by
Henry et al. (2010), and our measured $F$([Ne~{\sc ii}] 12.8 $\mu$m) 
=  6.83(--12) erg s$^{-1}$ cm$^{-2}$, we estimated 
C and Ne abundances. To do so, we assumed that C/H =
(Ar/Ar$^{2+}$)$\times$C$^{2+}$/H$^{+}$ and Ne/H = Ne$^{+}$/H$^{+}$.
We derived  C$^{2+}$/H$^{+}$ and Ne$^{+}$/H$^{+}$ as 2.06(--3) and 5.96(--5),
respectively. In the gas-phase abundances of all objects 
listed in Table \ref{c60}, the N, O, Ne, S, Cl, Ar, and Kr 
abundances are derived from CELs, and the He and C abundances from RLs. 
The CEL C abundances in M1-12 and M1-20 are unknown due to a lack
of UV spectra.

We found the following common properties in these PNe:
(1) relatively cool central stars except for M1-20, (2)
young age, (3) broad 6-9, 10-14, and 30 $\mu$m 
features, (4) a strong stellar wind. In the UV spectra taken by \emph{IUE} and 
$FUSE$, we found that Tc1 shows the P-Cygni 
profile of Ly$\alpha$ 1215 {\AA}. There are no data with enough 
signal-to-noise ratio to check for the presence of P-Cygni Ly$\alpha$ 
profiles in the other PNe. However, since these PNe have the 
Wolf-Rayet-like central stars ([WC10,11] and weak emission line central star, 
wel), which show broad C\,{\sc
iii,iv} lines ($\gtrsim$60 km s$^{-1}$ at FWHM), there would 
be a strong wind for each. 

From properties (1) and (2), the short time duration of weak radiation 
from the central star would be an essential condition for fullerenes to 
survive or to be observed for a long time. The observational fact that all
fullerene PNe have cool central stars and low-excitation
nebulae, except for M1-20, matches the detection
of the C$_{60}$ in a proto PN and two post AGB stars by Zhang \& Kwok
(2011) and Gielen et al. (2011), respectively; these sources have 
cool central stars (21\,500 K in IRAS01005+7910; 6250 and 6000 K in
IRAS06338+5333 and HD52961, respectively). 
Property (3) implies that fullerenes might not be formed from HACs in
Galactic PNe. HACs reproduce 6-9, 10-14, $\sim$20, and $\sim$30 $\mu$m broad features. If the 
30 $\mu$m feature is from HACs, we should see the $\sim$20 $\mu$m feature, 
too. However, not all Galactic fullerene PNe show the 
$\sim$20 $\mu$m feature. As we mentioned in Section 2.6.2, however, the strength of 20 and 30
$\mu$m features largely depends on H-content, grain size, etc. At
present, therefore, we cannot rule out that fullerenes arise from the destruction of HACs. From property (4), the strong stellar wind 
would promote fullerene synthesis for young PNe and post-AGB stars
with cool central stars. In M1-11, as we mentioned in Section 2.3,
the H$_{2}$ lines within the optical diameter are excited by both
UV fluorescence and shocks. Lumsden et
al. (2001) observed the $HK$-band spectra of M1-12 and M1-20. 
However H$_{2}$ 1-0S(1) and 2-1S(1) lines are not detected.

If C$_{60}$ is from large PAH clusters, the spatial distribution 
of PAHs and C$_{60}$ would be similar to that in NGC7023. 
If C$_{60}$ is from HACs, the spatial distribution of C$_{60}$ 
would be similar to that in Tc1, as shown in Fig.3 of 
Bernard-Salas et al. (2012). 
At present, destruction of HACs seems to be the most plausible 
scenario for fullerene formation in Galactic PNe, including M1-11. 
If so, the spatial distribution of C$_{60}$ in M1-11 
would be similar to that in Tc1. To advance our understanding of 
what kinds of conditions in which fullerenes can be formed and survive, the
high-resolution spatial observations are necessary to investigate the
spatial variation of HACs, PAHs, C$_{60}$, and dust components.

\section{Summary and Future work}
We performed multiwavelength observations for the young PN M1-11 to
investigate elemental abundances and dust mass, 
and we discussed its evolutionary  status from its progenitor. 
We found a large discrepancy between the RL and the CEL O \& N
abundances. The RL C/O abundance ratio is $<$1. 
If the RL C/O ratio represents the chemistry of the PN, 
the nebula is O-rich. The RL O abundance is
much larger than that predicted by AGB nucleosynthesis models, 
while the CEL O abundance is close to that predicted by 
the models. The CEL C and O
would be reliable in M1-11. Therefore, we estimated the CEL 
C abundance using the relation between ADF(C$^{2+}$) and 
ADF(O$^{2+}$) among 56 Galactic PNe, and we obtained a CEL C/O, 
implying the possibility that M1-11 could be C-rich. 
We estimated the CEL C and O abundances through the SED model.

In the \emph{AKARI}/IRC, VLT/VISIR, and \emph{Spitzer}/IRS
spectra, 3.3, 8.6, and 11.3 $\mu$m PAH bands and broad 10-13 and 30 $\mu$m 
features are visible. SiC and MgS could be the main contributors 
to the broad 10-13 and 30 $\mu$m features, respectively. We detect broad emission from $\sim$16-22 $\mu$m; however, it 
is still unclear whether this feature is from C-rich or O-rich dust. 
We also detect three C$_{60}$ lines in the VISIR and \emph{Spitzer} data
of M1-11. The presence of C$_{60}$ might be explained by destruction of
HACs in this PN. As shown in Table \ref{c60}, fullerene-containing 
Galactic PNe seem to have central
star properties, gas-phase abundances, and dust composition in common. Except for M1-20,
most of these PNe that have fullerenes seem to have cool central
stars. The radiation field around a relatively low temperature star is
intrinsically weak. Moreover, the surrounding shell structure around
such a low temperature star might have diluted the stellar UV flux. 
Such circumstances must have played a favorable role for fullerenes 
to survive in the circumstellar medium. 

Through $\sim$0.1-90 $\mu$m SED modeling, we estimate the dust mass
and the properties of the central star and the nebula. Our current
analysis indicates that the progenitor would be a 1-1.5 $M_{\odot}$
star. The observed abundances and the predicted CEL C and C/O 
can be explained by an AGB nucleosynthesis model for 1.5 $M_{\odot}$
progenitors with $Z$=0.004. M1-11 is possibly a C-rich PN.

For a future study on M1-11, high spatial
resolution spectroscopy or narrow-band imaging 
would be necessary to resolve the spatial 
distribution of PAHs, C$_{60}$, HAC, and dust components 
to check how these co-exist in the
nebulae. It is essential to investigate the properties 
of the central stars, the excitation
degree of the nebulae, and the dust composition in young and
low-excitation PNe such as M1-11 to understand the condition in which 
fullerenes form and survive. 

\section*{Acknowledgments}
We are grateful to the anonymous referee for a  careful review
and many valuable suggestions. The authors thank Prof. Karen
Kwitter for providing their M1-11 spectrum.
M.O. acknowledges funding support from STScI GO-1129.01-A, NASA
NAO-50-12595, and STScI DDRF D0101.90128. FK acknowledges 
support from the National Science Council in the form of 
grant NSC100-2112-M-001-023-MY3. S.H. acknowledges support by the
Basic Research Program through the National Research Foundation of Korea
by the Ministry of Education, Science and Technology
(NRF-2011-0005077).
This work is mainly based on data
collected at the Subaru Telescope, which is operated by
the National Astronomical Observatory of Japan (NAOJ).
This work is in part based on \emph{HST} and \emph{IUE} archive data
downloaded from the $MAST$ and \emph{AKARI} archive data from $DARTS$.
This work is in part based on archival data obtained
with the Spitzer Space Telescope, which is operated by the
Jet Propulsion Laboratory, California Institute of
Technology under a contract with NASA. Support for this
work was provided by an award issued by JPL/Caltech.

\appendix
\section*{Observed Line Lists}
The detected lines in the Subaru/HDS and OAO/ISLE are listed in Tables
16 and 17, respectively. 

\begin{deluxetable*}{@{}cl@{\hspace{3pt}}r@{\hspace{3pt}}r|cl@{\hspace{3pt}}r@{\hspace{3pt}}r|cl@{\hspace{3pt}}r@{\hspace{3pt}}r@{}}
\tablecolumns{12}
\tablecaption{The detected lines in the HDS observations.}
\tablewidth{480pt}
\tablehead{
\colhead{$\lambda_{\rm lab}$}&
\colhead{Ion}&
\colhead{$f$($\lambda$)}&
\colhead{$I$($\lambda$)}&
\colhead{$\lambda_{\rm lab}$}&
\colhead{Ion}&
\colhead{$f$($\lambda$)}&
\colhead{$I$($\lambda$)}&
\colhead{$\lambda_{\rm lab}$}&
\colhead{Ion}&
\colhead{$f$($\lambda$)}&
\colhead{$I$($\lambda$)}\\
\colhead{({\AA})}            &
\colhead{}&              &
\colhead{[$I$({\hb})=100]}&
\colhead{({\AA})}            &
\colhead{}&              &
\colhead{[$I$({\hb})=100]}&
\colhead{({\AA})}            &
\colhead{}&              &
\colhead{[$I$({\hb})=100]}
}
\startdata
3661.21	&	 H31 	&	0.335	&	 0.681 $\pm$ 0.111 	&	4574.88	&	 [Mn~{\sc iii}] 	&	0.083	 &	 0.059 $\pm$ 0.010 	&	5537.60	&	 [Cl~{\sc iii}] 	&	 --0.149 	&	 0.072 $\pm$ 0.012 \\
3662.26	&	 H30 	&	0.335	&	 0.698 $\pm$ 0.073 	&	4591.12	&	 S~{\sc ii} 	&	0.078	&	 0.082 $\pm$ 0.015 	&	5554.83	&	 O~{\sc i} 	&	 --0.152 	&	 0.100 $\pm$ 0.023 \\
3663.40	&	 H29 	&	0.335	&	 0.483 $\pm$ 0.062 	&	4630.54	&	 N~{\sc ii} 	&	0.066	&	 0.055 $\pm$ 0.007 	&	5577.20	&	 [O~{\sc i}] 	&	 --0.156 	&	 0.040 $\pm$ 0.008 \\
3664.68	&	 H28 	&	0.335	&	 0.865 $\pm$ 0.075 	&	4634.12	&	 N~{\sc iii} 	&	0.065	&	 0.447 $\pm$ 0.027 	&	5659.60	&	 C~{\sc iii} 	&	 --0.169 	&	 0.054 $\pm$ 0.011 \\
3666.10	&	 H27 	&	0.334	&	 0.465 $\pm$ 0.063 	&	4635.32	&	 Fe~{\sc ii} 	&	0.065	&	 0.033 $\pm$ 0.004 	&	5666.63	&	 N~{\sc ii} 	&	 --0.171 	&	 0.101 $\pm$ 0.048 \\
3667.71	&	 H26 	&	0.334	&	 0.583 $\pm$ 0.079 	&	4640.03	&	 C~{\sc iii} 	&	0.063	&	 0.012 $\pm$ 0.001 	&	5690.43	&	 Si~{\sc i} 	&	 --0.174 	&	 0.038 $\pm$ 0.017 \\
3669.46	&	 H25 	&	0.334	&	 1.135 $\pm$ 0.082 	&	4640.64	&	 N~{\sc iii} 	&	0.063	&	 0.115 $\pm$ 0.004 	&	5695.92	&	 C~{\sc iii} 	&	 --0.175 	&	 2.113 $\pm$ 0.036 \\
3671.38	&	 H24 	&	0.333	&	 0.961 $\pm$ 0.063 	&	4641.81	&	 O~{\sc ii} 	&	0.063	&	 0.026 $\pm$ 0.001 	&	5730.66	&	 N~{\sc ii} 	&	 --0.181 	&	 0.043 $\pm$ 0.013 \\
3673.74	&	 H23 	&	0.333	&	 1.182 $\pm$ 0.075 	&	4641.85	&	 N~{\sc iii} 	&	0.063	&	 0.044 $\pm$ 0.002 	&	5754.60	&	 [N~{\sc ii}] 	&	 --0.185 	&	 5.986 $\pm$ 0.077 \\
3674.84	&	 He~{\sc ii} 	&	0.333	&	 0.174 $\pm$ 0.045 	&	4649.14	&	 O~{\sc ii} 	&	 0.060	 &	 0.022 $\pm$ 0.006 	&	5826.42	&	 C~{\sc iii} 	&	 --0.196 	&	 0.122 $\pm$ 0.024 \\
3676.36	&	 H22 	&	0.332	&	 1.025 $\pm$ 0.076 	&	4650.84	&	 O~{\sc ii} 	&	0.060	&	 0.031 $\pm$ 0.007 	&	5831.70	&	 N~{\sc iii}? 	&	 --0.196 	&	 0.041 $\pm$ 0.012 \\
3679.35	&	 H21 	&	0.332	&	 1.384 $\pm$ 0.074 	&	4652.05	&	 C~{\sc iii} 	&	0.060	&	 0.213 $\pm$ 0.019 	&	5875.66	&	 He~{\sc i} 	&	 --0.203 	&	 6.155 $\pm$ 0.083 \\
3682.81	&	 H20 	&	0.331	&	 1.482 $\pm$ 0.058 	&	4658.64	&	 C~{\sc iv} 	&	0.058	&	 0.137 $\pm$ 0.008 	&	5958.39	&	 O~{\sc i} 	&	 --0.215 	&	 0.129 $\pm$ 0.021 \\
3686.83	&	 H19 	&	0.330	&	 1.569 $\pm$ 0.054 	&	4676.23	&	 O~{\sc ii} 	&	0.053	&	 0.096 $\pm$ 0.023 	&	6046.23	&	 O~{\sc i} 	&	 --0.228 	&	 0.239 $\pm$ 0.028 \\
3691.55	&	 H18 	&	0.329	&	 2.015 $\pm$ 0.097 	&	4685.68	&	 He~{\sc ii} 	&	0.050	&	 0.078 $\pm$ 0.021 	&	6074.20	&	 C~{\sc i} 	&	 --0.232 	&	 0.057 $\pm$ 0.014 \\
3697.15	&	 H17 	&	0.328	&	 1.853 $\pm$ 0.065 	&	4701.53	&	 [Fe~{\sc iii}] 	&	0.045	 &	 0.069 $\pm$ 0.008 	&	6077.90	&	 C~{\sc ii} 	&	 --0.232 	&	 0.075 $\pm$ 0.021 \\
3703.65	&	 H16 	&	0.327	&	 1.741 $\pm$ 0.051 	&	4713.17	&	 He~{\sc i} 	&	0.042	&	 0.260 $\pm$ 0.012 	&	6267.81	&	 [V~{\sc ii}]? 	&	 --0.258 	&	 0.041 $\pm$ 0.010 \\
3711.97	&	 H15 	&	0.325	&	 2.328 $\pm$ 0.052 	&	4754.69	&	 [Fe~{\sc iii}] 	&	0.030	 &	 0.039 $\pm$ 0.009 	&	6300.34	&	 [O~{\sc i}] 	&	 --0.263 	&	 2.203 $\pm$ 0.041 \\
3715.08	&	 O~{\sc iii} 	&	0.324	&	 0.296 $\pm$ 0.041 	&	4769.43	&	 [Fe~{\sc iii}] 	&	 0.025	&	 0.053 $\pm$ 0.006 	&	6312.10	&	 [S~{\sc iii}] 	 &	 --0.264 	&	 0.323 $\pm$ 0.015 \\
3721.94	&	 H14 	&	0.323	&	 2.583 $\pm$ 0.074 	&	4789.57	&	 N~{\sc ii} 	&	0.017	&	 0.051 $\pm$ 0.016 	&	6347.09	&	 Si~{\sc ii} 	&	 --0.269 	&	 0.126 $\pm$ 0.017 \\
3726.03	&	 [O~{\sc ii}] 	&	0.322	&	 55.380 $\pm$ 0.526 	&	4861.33	&	 H4 	&	0.000	 &	 100.000 $\pm$ 0.130 	&	6363.78	&	 [O~{\sc i}] 	&	 --0.271 	&	 0.748 $\pm$ 0.021 \\
3728.81	&	 [O~{\sc ii}] 	&	0.322	&	 19.692 $\pm$ 0.302 	&	4881.00	&	 [Fe~{\sc iii}] 	 &	 --0.005 	&	 0.049 $\pm$ 0.005 	&	6527.24	&	 [N~{\sc ii}] 	&	 --0.293 	&	 0.030 $\pm$ 0.004 \\
3734.37	&	 H13 	&	0.321	&	 2.590 $\pm$ 0.064 	&	4921.93	&	 He~{\sc i} 	&	 --0.016 	 &	 0.464 $\pm$ 0.007 	&	6548.10	&	 [N~{\sc ii}] 	&	 --0.296 	&	 58.851 $\pm$ 0.989 \\
3750.15	&	 H12 	&	0.317	&	 3.438 $\pm$ 0.051 	&	4924.54	&	 [Fe~{\sc iii}] 	&	 --0.017 	 &	 0.016 $\pm$ 0.008 	&	6562.77	&	 H3	&	 --0.298 	 &	 285.000 $\pm$ 4.160 \\
3770.63	&	 H11 	&	0.313	&	 3.833 $\pm$ 0.069 	&	4931.80	&	 [O~{\sc iii}] 	&	 --0.019 	 &	 0.136 $\pm$ 0.026 	&	6578.01	&	 C~{\sc ii} 	&	 --0.300 	&	 0.499 $\pm$ 0.012 \\
3797.90	&	 H10 	&	0.307	&	 7.047 $\pm$ 0.065 	&	4934.08	&	 Ba~{\sc ii} 	&	 --0.019 	 &	 0.016 $\pm$ 0.004 	&	6583.50	&	 [N~{\sc ii}] 	&	 --0.300 	&	 190.199 $\pm$ 4.767 \\
3819.60	&	 He~{\sc i} 	&	0.302	&	 0.363 $\pm$ 0.034 	&	4958.91	&	 [O~{\sc iii}] 	&	 --0.026 	&	 6.440 $\pm$ 0.031 	&	6678.16	&	 He~{\sc i} 	&	 --0.313 	&	 1.600 $\pm$ 0.028 \\
3835.38	&	 H9 	&	0.299	&	 8.950 $\pm$ 0.083 	&	5006.84	&	 [O~{\sc iii}] 	&	 --0.038 	 &	 18.544 $\pm$ 0.082 	&	6688.79	&	 C~{\sc i} 	&	 --0.314 	&	 0.050 $\pm$ 0.012 \\
3889.05	&	 H8 	&	0.286	&	 13.681 $\pm$ 0.100 	&	5015.68	&	 He~{\sc i} 	&	 --0.040 	 &	 0.989 $\pm$ 0.004 	&	6701.49	&	 N~{\sc ii} 	 &	 --0.316 	&	 0.067 $\pm$ 0.008 \\
3918.97	&	 C~{\sc ii} 	&	0.279	&	 0.242 $\pm$ 0.032 	&	5017.22	&	 N~{\sc ii} 	&	 --0.040 	&	 0.014 $\pm$ 0.004 	&	6716.44	&	 [S~{\sc ii}] 	 &	 --0.318 	&	 0.442 $\pm$ 0.016 \\
3920.68	&	 C~{\sc ii} 	&	0.279	&	 0.510 $\pm$ 0.025 	&	5030.33	&	 [Fe~{\sc iv}] 	&	 --0.043 	&	 0.036 $\pm$ 0.014 	&	6730.82	&	 [S~{\sc ii}] 	&	 --0.320 	&	 1.084 $\pm$ 0.027 \\
3964.73	&	 He~{\sc i} 	&	0.267	&	 0.598 $\pm$ 0.023 	&	5032.07	&	 C~{\sc ii} 	&	 --0.044 	&	 0.106 $\pm$ 0.018 	&	6744.33	&	 C~{\sc iii} 	 &	 --0.322 	&	 0.125 $\pm$ 0.017 \\
3970.07	&	 H7 	&	0.266	&	 18.792 $\pm$ 0.104 	&	5041.02	&	 Si~{\sc ii} 	&	 --0.046 	 &	 0.043 $\pm$ 0.008 	&	7002.17	&	 O~{\sc i} 	 &	 --0.356 	&	 0.210 $\pm$ 0.009 \\
4026.18	&	 He~{\sc i} 	&	0.251	&	 0.949 $\pm$ 0.020 	&	5047.74	&	 He~{\sc i} 	&	 --0.048 	&	 0.105 $\pm$ 0.006 	&	7018.63	&	 Ca~{\sc i}]? 	 &	 --0.358 	&	 0.023 $\pm$ 0.006 \\
4056.91	&	 N~{\sc ii} 	&	0.243	&	 0.254 $\pm$ 0.029 	&	5055.98	&	 Si~{\sc ii} 	&	 --0.050 	&	 0.119 $\pm$ 0.011 	&	7037.25	&	 C~{\sc iii} 	&	 --0.361 	&	 0.139 $\pm$ 0.013 \\
4068.60	&	 [S~{\sc ii}] 	&	0.239	&	 1.790 $\pm$ 0.047 	&	5080.49	&	 [Ni~{\sc iii}] 	&	 --0.056 	 &	 0.137 $\pm$ 0.023 	&	7065.18	&	 He~{\sc i} 	&	 --0.364 	&	 2.276 $\pm$ 0.036 \\
4076.35	&	 [S~{\sc ii}] 	&	0.237	&	 0.634 $\pm$ 0.055 	&	5121.83	&	 C~{\sc ii} 	&	 --0.065 	&	 0.063 $\pm$ 0.009 	&	7065.71	&	 He~{\sc i} 	 &	 --0.364 	&	 0.334 $\pm$ 0.011 \\
4089.30	&	 O~{\sc iii} 	&	0.233	&	 0.095 $\pm$ 0.014 	&	5131.25	&	 [Kr{\sc v}] 	&	 --0.067 	&	 0.140 $\pm$ 0.015 	&	7101.07	&	 Ca~{\sc i} 	&	 --0.369 	&	 0.020 $\pm$ 0.006 \\
4101.73	&	 H6 	&	0.230	&	 27.822 $\pm$ 0.127 	&	5143.29	&	 [Fe~{\sc iii}] 	&	 --0.070 	&	 0.048 $\pm$ 0.003 	&	7135.80	&	 [Ar~{\sc iii}] 	&	 --0.374 	&	 2.457 $\pm$ 0.042 \\
4128.66	&	 N~{\sc ii} 	&	0.222	&	 0.216 $\pm$ 0.052 	&	5145.17	&	 C~{\sc ii} 	&	 --0.070 	&	 0.092 $\pm$ 0.005 	&	7179.88	&	 Ca~{\sc i} 	 &	 --0.380 	&	 0.066 $\pm$ 0.006 \\
4143.76	&	 He~{\sc i} 	&	0.217	&	 0.175 $\pm$ 0.019 	&	5145.75	&	 [Fe~{\sc vi}]? 	&	 --0.071 	 &	 0.018 $\pm$ 0.008 	&	7231.33	&	 C~{\sc ii} 	&	 --0.387 	&	 0.216 $\pm$ 0.012 \\
4267.15	&	 C~{\sc ii} 	&	0.180	&	 0.614 $\pm$ 0.027 	&	5146.45	&	 [Fe~{\sc iii}] 	&	 --0.071 	 &	 0.115 $\pm$ 0.007 	&	7231.48	&	 Ca~{\sc i} 	&	 --0.387 	&	 0.077 $\pm$ 0.007 \\
4307.44	&	 N~{\sc iii}? 	&	0.167	&	 0.099 $\pm$ 0.019 	&	5191.82	&	 [Ar~{\sc iii}] 	&	 --0.081 	 &	 0.019 $\pm$ 0.004 	&	7236.20	&	 Ca~{\sc i} 	&	 --0.387 	&	 0.058 $\pm$ 0.004 \\
4317.19	&	 C~{\sc iii} 	&	0.164	&	 0.120 $\pm$ 0.018 	&	5197.90	&	 [N~{\sc i}] 	&	 --0.082 	&	 0.201 $\pm$ 0.009 	&	7236.42	&	 C~{\sc ii} 	&	 --0.387 	&	 0.436 $\pm$ 0.011 \\
4340.46	&	 H5 	&	0.157	&	 46.900 $\pm$ 0.174 	&	5200.26	&	 [N~{\sc i}] 	&	 --0.083 	 &	 0.123 $\pm$ 0.008 	&	7236.80	&	 C~{\sc i}] 	&	 --0.387 	&	 0.057 $\pm$ 0.004 \\
4352.52	&	 N~{\sc ii} 	&	0.153	&	 0.028 $\pm$ 0.011 	&	5270.40	&	 [Fe~{\sc iii}] 	&	 --0.098 	 &	 0.071 $\pm$ 0.004 	&	7237.17	&	 C~{\sc ii} 	&	 --0.387 	&	 0.071 $\pm$ 0.005 \\
4363.21	&	 [O~{\sc iii}] 	&	0.149	&	 0.137 $\pm$ 0.015 	&	5274.78	&	 [Mn~{\sc ii}] 	&	 --0.099 	&	 0.055 $\pm$ 0.009 	&	7254.53	&	 O~{\sc i} 	 &	 --0.390 	&	 0.263 $\pm$ 0.012 \\
4368.24	&	 O~{\sc i} 	&	0.148	&	 0.317 $\pm$ 0.029 	&	5298.88	&	 [Fe~{\sc ii}] 	&	 --0.104 	 &	 0.118 $\pm$ 0.009 	&	7281.35	&	 He~{\sc i} 	&	 --0.393 	&	 0.319 $\pm$ 0.013 \\
4369.56	&	 O~{\sc iii} 	&	0.147	&	 0.137 $\pm$ 0.031 	&	5304.54	&	 C~{\sc iii} 	&	 --0.105 	&	 0.063 $\pm$ 0.016 	&	7318.92	&	 [O~{\sc ii}] 	&	 --0.398 	&	 3.837 $\pm$ 0.077 \\
4377.77	&	 Mn~{\sc ii} 	&	0.145	&	 0.060 $\pm$ 0.016 	&	5342.43	&	 C~{\sc ii} 	&	 --0.112 	&	 0.032 $\pm$ 0.004 	&	7319.99	&	 [O~{\sc ii}] 	 &	 --0.398 	&	 11.165 $\pm$ 0.193 \\
4387.93	&	 He~{\sc i} 	&	0.142	&	 0.238 $\pm$ 0.008 	&	5347.89	&	 O~{\sc ii} 	&	 --0.113 	&	 0.018 $\pm$ 0.005 	&	7329.67	&	 [O~{\sc ii}] 	 &	 --0.400 	&	 5.513 $\pm$ 0.103 \\
4414.90	&	 O~{\sc ii} 	&	0.133	&	 0.191 $\pm$ 0.017 	&	5351.81	&	 [Cr~{\sc iii}] 	&	 --0.114 	 &	 0.047 $\pm$ 0.006 	&	7330.73	&	 [O~{\sc ii}] 	&	 --0.400 	&	 5.284 $\pm$ 0.102 \\
4416.98	&	 O~{\sc ii} 	&	0.132	&	 0.223 $\pm$ 0.020 	&	5365.10	&	 [Mn~{\sc iii}] 	&	 --0.117 	 &	 0.022 $\pm$ 0.004 	&	7442.20	&	 C~{\sc iv} 	&	 --0.415 	&	 0.076 $\pm$ 0.007 \\
4471.47	&	 He~{\sc i} 	&	0.115	&	 1.718 $\pm$ 0.030 	&	5461.91	&	 Fe~{\sc ii} 	&	 --0.135 	&	 0.038 $\pm$ 0.021 	&	7452.60	&	 [Fe~{\sc ii}] 	&	 --0.416 	&	 0.119 $\pm$ 0.010 \\
4541.59	&	 He~{\sc ii} 	&	0.094	&	 0.249 $\pm$ 0.021 	&	5494.69	&	 [Fe~{\sc iii}] 	&	 --0.141 	 &	 0.064 $\pm$ 0.014 	&	7457.90	&	 C~{\sc ii}] 	&	 --0.417 	&	 0.044 $\pm$ 0.005 \\
4571.10	&	 Mg~{\sc i}] 	&	0.084	&	 0.221 $\pm$ 0.017 	&	5517.66	&	 [Cl~{\sc iii}] 	&	 --0.145 	 &	 0.029 $\pm$ 0.006 	&	7468.31	&	 N~{\sc i} 	 &	 --0.418 	&	 0.103 $\pm$ 0.007
\enddata
\label{hdst}
\end{deluxetable*}

\begin{deluxetable*}{@{}l@{\hspace{2pt}}lcc@{\hspace{-4pt}}r@{}}
\tablecolumns{5}
\tablecaption{The detected lines in the ISLE observations.}
\tablewidth{480pt}
\tablehead{
\colhead{$\lambda_{vac}$}&\colhead{Ion}&\colhead{$f$($\lambda$)}&\colhead{$F(\lambda)$}&\colhead{$I(\lambda)$}\\
\colhead{($\mu$m)}& \colhead{}& \colhead{}&\colhead{(erg s $^{-1}$ cm$^{-2}$)}&\colhead{($I$(Pa$\beta$,Br$\gamma$)=100)}
}
\startdata
1.129 & O \,{\sc i} & --0.715 & 7.34(--14) $\pm$ 4.51(--15) & 2.535 $\pm$ 0.188 \\ 
1.147 & [P \,{\sc ii}] & --0.722 & 2.23(--14) $\pm$ 3.20(--15) & 0.755 $\pm$ 0.113 \\ 
1.162 & H$_{2}$ 2-0 S(1) & --0.728 &8.51(--15) $\pm$ 9.80(--16) & 0.283 $\pm$ 0.035 \\ 
1.175 & O\,{\sc i}? & --0.733 & 1.76(--14) $\pm$ 1.29(--15) & 0.577 $\pm$ 0.049 \\ 
1.186 & H$_{2}$ 3-1 S(3) & --0.736 & 1.62(--14) $\pm$ 9.60(--16) & 0.525 $\pm$ 0.038 \\ 
1.189 & [P\,{\sc ii}] & --0.737 & 6.50(--14) $\pm$ 1.06(--15) & 2.106 $\pm$ 0.095 \\ 
1.190 & H$_{2}$ 2-0 S(0) & --0.738 & 6.46(--15) $\pm$ 1.06(--15) & 0.209 $\pm$ 0.035 \\ 
1.194 & Ca\,{\sc i}? & --0.739 & 5.12(--15) $\pm$ 1.37(--15) & 0.165 $\pm$ 0.045 \\ 
1.197 & He \,{\sc i} & --0.740 & 8.09(--15) $\pm$ 8.40(--16) & 0.260 $\pm$ 0.029 \\ 
1.199 & N\,{\sc iii}? & --0.741 & 7.32(--15) $\pm$ 6.80(--16) & 0.235 $\pm$ 0.024 \\ 
1.208 & H$_{2}$ 3-1 S(2) & --0.744 &8.43(--15) $\pm$ 7.90(--16) & 0.268 $\pm$ 0.028 \\ 
1.219 & C\,{\sc i}? & --0.748 & 4.69(--15) $\pm$ 7.90(--16) & 0.147 $\pm$ 0.026 \\ 
1.221 & Ca \,{\sc i}? & --0.748 & 3.44(--15) $\pm$ 7.90(--16) & 0.108 $\pm$ 0.025 \\ 
1.226 & H$_{2}$ 4-2 S(5) & --0.750 & 4.79(--15) $\pm$ 6.80(--16) & 0.150 $\pm$ 0.022 \\ 
1.229 & C\,{\sc ii} & --0.751 & 1.05(--14) $\pm$ 6.60(--16) & 0.327 $\pm$ 0.025 \\ 
1.230 & Si\,{\sc i}? & --0.751 & 7.11(--15) $\pm$ 6.60(--16) & 0.221 $\pm$ 0.023 \\ 
1.233 & H$_{2}$ 3-1S(1) & --0.752 & 2.07(--14) $\pm$ 8.80(--16) & 0.642 $\pm$ 0.039 \\ 
1.239 & H$_{2}$ 2-0 Q(1) & --0.754 & 1.50(--14) $\pm$ 1.08(--15) & 0.462 $\pm$ 0.039 \\ 
1.242 & H$_{2}$ 4-2 S(4) & --0.755 & 7.44(--15) $\pm$ 1.05(--15) & 0.229 $\pm$ 0.034 \\ 
      & + H$_{2}$ 2-0Q(2)\\
1.247 & Si\,{\sc i}? & --0.757 & 4.84(--15) $\pm$ 2.08(--15) & 0.149 $\pm$ 0.064 \\ 
1.247 & H$_{2}$ 2-0 Q(3) & --0.757 & 8.61(--15) $\pm$ 8.50(--16) & 0.264 $\pm$ 0.028 \\ 
1.253 & He \,{\sc i} & --0.759 & 2.05(--14) $\pm$ 9.90(--16) & 0.625 $\pm$ 0.040 \\ 
1.255 & H$_{2}$ 2-0 Q(4) & --0.759 & 2.25(--15) $\pm$ 8.70(--16) & 0.069 $\pm$ 0.027 \\ 
1.257 & [Fe \,{\sc ii}] & --0.760 & 6.52(--15) $\pm$ 7.50(--16) & 0.198 $\pm$ 0.024 \\ 
1.262 & H$_{2}$ 4-2 S(3) & --0.761 & 1.25(--14) $\pm$ 8.90(--16) & 0.379 $\pm$ 0.031 \\ 
1.262 & H$_{2}$ 3-1S(0) & --0.761 & 9.17(--15) $\pm$ 8.90(--16) & 0.278 $\pm$ 0.029 \\ 
1.264 & H$_{2}$ 2-0Q(5) & --0.762 & 3.23(--15) $\pm$ 1.15(--15) & 0.098 $\pm$ 0.035 \\ 
1.265 & [Fe\,{\sc iv}]? & --0.762 & 4.18(--15) $\pm$ 9.10(--16) & 0.126 $\pm$ 0.028 \\ 
1.2788 & He \,{\sc i} & --0.766 & 4.16(--14) $\pm$ 1.31(--15) & 1.242 $\pm$ 0.066 \\ 
1.2794 & He \,{\sc i} & --0.767 & 1.92(--14) $\pm$ 1.31(--15) & 0.573 $\pm$ 0.046 \\ 
1.282 & Pa$\beta$ & --0.767 & 3.36(--12) $\pm$ 3.05(--15) & 100.000 $\pm$ 4.290 \\ 
1.285 & H$_{2}$ 4-2 S(2) & --0.768 &8.42(--15) $\pm$ 8.30(--16) & 0.250 $\pm$ 0.027 \\ 
2.034 & H$_{2}$ 1-0 S(2) & --0.889 &1.97(--14) $\pm$ 1.48(--15) & 3.476 $\pm$ 0.315 \\ 
2.042 & H$_{2}$ 8-6 O(3)  & --0.890 &1.43(--14) $\pm$ 1.63(--15) & 2.519 $\pm$ 0.314 \\ 
2.059 & He \,{\sc i} & --0.891 & 2.74(--13) $\pm$ 3.83(--15) & 48.053 $\pm$ 2.534 \\ 
2.074 & H$_{2}$ 2-1 S(3) & --0.893 & 2.15(--14) $\pm$ 1.33(--15) & 3.758 $\pm$ 0.301 \\ 
2.102 & Si \,{\sc i}?  & --0.895 & 9.27(--15) $\pm$ 9.50(--16) & 1.610 $\pm$ 0.184 \\ 
2.113 & He \,{\sc i} & --0.896 & 7.77(--15) $\pm$ 1.27(--15) & 1.346 $\pm$ 0.230 \\ 
2.122 & H$_{2}$ 1-0 S(1) & --0.897 & 4.90(--14) $\pm$ 1.05(--15) & 8.471 $\pm$ 0.468 \\ 
2.138 & Mg\,{\sc ii} & --0.898 & 4.67(--15) $\pm$ 5.30(--16) & 0.805 $\pm$ 0.100 \\ 
2.144 & Mg\,{\sc ii} & --0.898 & 4.88(--15) $\pm$ 8.40(--16) & 0.840 $\pm$ 0.151 \\ 
2.154 & H$_{2}$ 2-1 S(2) & --0.899 & 8.00(--15) $\pm$ 8.10(--16) & 1.373 $\pm$ 0.156 \\ 
2.162 & He \,{\sc i} & --0.900 & 9.78(--15) $\pm$ 3.30(--16) & 1.676 $\pm$ 0.103 \\ 
2.166 & H \,{\sc i} (Br$\gamma$) & -0.900 & 5.84(--13) $\pm$ 5.23(--15) & 100.000 $\pm$ 5.186 \\ 
2.171 & O\,{\sc i} & --0.900 & 8.54(--15) $\pm$ 1.08(--15) & 1.461 $\pm$ 0.199 \\ 
2.189 & He \,{\sc ii} & --0.902 & 6.50(--15) $\pm$ 7.60(--16) & 1.108 $\pm$ 0.141 \\ 
2.199 & [Kr \,{\sc iii}] & --0.902 & 1.89(--14) $\pm$ 1.22(--15) & 3.215 $\pm$ 0.265 \\ 
2.201 & H$_{2}$ 3-2 S(3) & --0.903 & 6.70(--15) $\pm$ 1.22(--15) & 1.139 $\pm$ 0.215 \\ 
2.205 & C\,{\sc ii}? & --0.903 & 2.82(--15) $\pm$ 7.40(--16) & 0.479 $\pm$ 0.128 \\ 
2.224 & H$_{2}$ 1-0 S(0) & --0.904 & 1.07(--14) $\pm$ 1.11(--15) & 1.811 $\pm$ 0.210 \\ 
2.248 & H$_{2}$ 2-1 S(1) & --0.906 & 1.99(--14) $\pm$ 1.20(--15) & 3.353 $\pm$ 0.265 \\ 
2.287 & H$_{2}$ 3-2 S(2) & --0.908 & 4.00(--15) $\pm$ 1.80(--15) & 0.669 $\pm$
 0.303 
\enddata
\label{isle_spec}
\end{deluxetable*}

\clearpage

\end{document}